\documentclass[prx,twocolumn,floatfix,superscriptaddress,longbibliography,nofootinbib]{revtex4-1}

\usepackage{amssymb,amsmath,amstext,amsthm,mathtools}
\usepackage{graphicx}
\usepackage{epstopdf}
\usepackage{color}
\usepackage{bm}
\usepackage{appendix}
\usepackage[T1]{fontenc}
\usepackage{bbold}
\usepackage{bbm}
\usepackage{latexsym}
\usepackage[colorlinks=true,citecolor=blue,linkcolor=magenta]{hyperref}
\usepackage{float}
\usepackage{verbatim}
\usepackage{lipsum}
\usepackage{braket}

\usepackage{tikz}
\usetikzlibrary{quantikz}

\newcommand{\dya}[1]{\ket{#1}\!\bra{#1}}

\newcommand{\bigzero}{\mbox{\normalfont\Large 0}}

\newcommand{\SC}{\mathcal{S}}

\newcommand{\Tr}{{\rm Tr}}

\renewcommand{\geq}{\geqslant}
\renewcommand{\leq}{\leqslant}

\renewcommand{\vec}[1]{\boldsymbol{#1}}  

\newcommand*{\id}{\openone}

\newcommand{\thv}{\vec{\theta}}
\newcommand{\gav}{\vec{\gamma}}
\newcommand{\alv}{\vec{\alpha}}

\newcommand{\gamv}{\vec{\gamma}}
\newcommand{\opt}{\rm opt}

\newtheoremstyle{example}{\topsep}{\topsep}%
{}
{}
{\bfseries}
{:}
{   }
{\thmname{#1}\thmnumber{ #2}}
\theoremstyle{example}

\theoremstyle{definition}

\begin{document}

\title{Long-time simulations with high fidelity on quantum hardware}

\author{Joe Gibbs}
\affiliation{
Theoretical Division, Los Alamos National Laboratory, Los Alamos, NM, USA.}
\author{Kaitlin Gili}
\affiliation{
Theoretical Division, Los Alamos National Laboratory, Los Alamos, NM, USA.}
\affiliation{
Department of Physics, University of Oxford, Clarendon Laboratory, Oxford, U.K.}
\author{Zo\"{e} Holmes}
\affiliation{
Information Sciences, Los Alamos National Laboratory, Los Alamos, NM, USA.}
\author{Benjamin Commeau}
\affiliation{
Information Sciences, Los Alamos National Laboratory, Los Alamos, NM, USA.}
\affiliation{
Department of Physics, University of Connecticut, Storrs, Connecticut, CT, USA.}
\author{Andrew Arrasmith} 
\affiliation{
Theoretical Division, Los Alamos National Laboratory, Los Alamos, NM, USA.}
\author{Lukasz Cincio}
\affiliation{
Theoretical Division, Los Alamos National Laboratory, Los Alamos, NM, USA.}
\author{Patrick J. Coles}
\affiliation{
Theoretical Division, Los Alamos National Laboratory, Los Alamos, NM, USA.}
\author{Andrew Sornborger}
\affiliation{
Information Sciences, Los Alamos National Laboratory, Los Alamos, NM, USA.}
\date{\today}

\begin{abstract}

Moderate-size quantum computers are now publicly accessible over the cloud, opening the exciting possibility of performing dynamical simulations of quantum systems. However, while rapidly improving, these devices have short coherence times, limiting the depth of algorithms that may be successfully implemented. Here we demonstrate that, despite these limitations, it is possible to implement long-time, high fidelity simulations on current hardware. Specifically, we simulate an XY-model spin chain on the Rigetti and IBM quantum computers, maintaining a fidelity of at least 0.9 for over 600 time steps. This is a factor of 150 longer than is possible using the iterated Trotter method. Our simulations are performed using a new algorithm that we call the fixed state Variational Fast Forwarding (fsVFF) algorithm. This algorithm decreases the circuit depth and width required for a quantum simulation by finding an approximate diagonalization of a short time evolution unitary. Crucially, fsVFF only requires finding a diagonalization on the subspace spanned by the initial state, rather than on the total Hilbert space as with previous methods, substantially reducing the required resources. We further demonstrate the viability of fsVFF through large numerical implementations of the algorithm, as well as an analysis of its noise resilience and the scaling of simulation errors.
\end{abstract}
\maketitle

\section{Introduction}\label{sc:intro}


The simulation of physical systems is both valuable for basic science and technological applications across a diverse range of industries, from materials design to pharmaceutical development. Relative to classical computers, quantum computers have the potential to provide an exponentially more efficient means of simulating quantum mechanical systems. Quantum hardware has progressed substantially in recent years~\cite{google2019supremacy, google2020observation}. However, despite continual progress, we remain in the `noisy intermediate-scale quantum' (NISQ) era in which the available hardware is limited to relatively small numbers of qubits and prone to errors. Simulation algorithms designed for fault-tolerant quantum computers, such as Trotterization methods~\cite{lloyd1996universal,sornborger1999higher}, qubitization methods~\cite{low2019hamiltonian}, and Taylor series methods~\cite{berry2015simulating}, require deeper circuits than viable given the short coherence times of current hardware.  Thus alternative approaches are needed to successfully implement useful simulations on NISQ hardware.





Variational quantum algorithms
~\cite{cerezo2020variationalreview,endo2021hybrid,bharti2021noisy,peruzzo2014variational,farhi2014quantum,mcclean2016theory,khatri2019quantum,larose2019variational,arrasmith2019variational,cerezo2020variationalfidelity,li2017efficient,endo2020variational,yao2020adaptive,heya2019subspace,cirstoiu2020variational,commeau2020variational}, where a classical computer optimizes a cost function measured on a quantum computer, show promise for NISQ quantum simulations. An early approach introduced an iterative method, where the state is variationally learned on a step-by-step basis using action principles~\cite{trout2018simulating, endo2020variational,yao2020adaptive, benedetti2020hardware}.
Subsequently, a generalization of the variational quantum eigensolver~\cite{peruzzo2014variational} was developed for simulations in low lying energy subspaces~\cite{heya2019subspace}. 
Very recently, quantum-assisted methods have been proposed that perform all necessary quantum measurements at the start of the algorithm instead of employing a classical-quantum feedback loop~\cite{bharti2020quantum,lau2021quantum,haug2020generalized}. 


In this work, we improve upon a recently proposed variational quantum algorithm known as Variational Fast Forwarding (VFF)~\cite{cirstoiu2020variational}. VFF allows long time simulations to be performed using a fixed depth circuit, thus enabling a quantum simulation to be `fast forwarded' beyond the coherence time of noisy hardware. The VFF algorithm requires finding a full diagonalization of the short time evolution operator $U$ of the system of interest. Once found, the diagonalization enables any initial state of that system to be fast forwarded. However, for practical purposes, one is often interested in studying the evolution of a particular fixed initial state of interest. In that case a full diagonalization of $U$ is overkill. Instead, it suffices to find a diagonal compilation of $U$ that captures its action on the given initial state. Here, we show that focusing on this commonly encountered but less exacting task can substantially reduce the resources required for the simulation.





Specifically, we introduce the fixed state VFF algorithm (fsVFF) for fast forwarding a fixed initial state beyond the coherence time of a quantum computer. This approach is tailored to making dynamical simulation more suitable for NISQ hardware in two key ways. First, the cost function requires half as many qubits as VFF. This not only allows larger scale simulations to be performed on current resource-limited hardware, but also has the potential to enable higher fidelity simulations since larger devices tend to be noisier. Second, fsVFF can utilize simpler ans\"{a}tze than VFF both in terms of the depth of the ansatz and the number of parameters that need to be learnt. Thus, fsVFF can reduce the width, depth, and total number of circuits required to fast forward quantum simulations, hence increasing the viability of performing simulations on near-term hardware. 


We demonstrate these advantages by implementing long-time high fidelity quantum simulations of the 2-qubit XY spin chain on Rigetti's and IBM's quantum computers. Specifically, while the iterated Trotter approach has a fidelity of less than 0.9 after 4 time steps and has completely thermalized by 25 time steps, with fsVFF we achieve a simulation fidelity greater than 0.9 for over 600 time steps. 
We further support the effectiveness of this approach for NISQ simulations, with 4 qubit noisy and 8 qubit noiseless numerical simulations of the XY model and Fermi-Hubbard model respectively. 

In our analytical results, we prove the faithfulness of the fsVFF cost function by utilizing the newly developed No-Free-Lunch theorems for quantum machine learning~\cite{poland2020no, sharma2020reformulation}. We also provide a proof of the noise resilience of the fsVFF cost function, specifically the optimal parameter resilience~\cite{sharma2019noise}. Finally, we perform an analysis of  simulation errors under fast-forwarding.

The diagonalizations obtained using fsVFF may further be useful for determining the eigenstates and eigenvalues of the Hamiltonian on the subspace spanned by the initial state. This can be done using a time series analysis, by using fsVFF to reduce the depth of the quantum phase estimation (QPE) algorithm, or using a simple sampling method. We demonstrate on IBM's quantum computer that, while standard QPE fails on real hardware, fsVFF can be used to obtain accurate estimates of the spectrum.

\section{Background}

Before presenting our fsVFF algorithm, let us first review the original VFF algorithm from Ref.~\cite{cirstoiu2020variational}. Consider a Hamiltonian $H$ on a $d=2^n$ dimensional Hilbert space (i.e., on $n$ qubits) evolved for a short time $\Delta t$ with the simulation unitary $e^{-iH\Delta t}$, and let $T$ (larger than $\Delta t$) denote the desired simulation time. Then the VFF algorithm consists of the following steps:


\begin{enumerate}
\item Approximate $e^{-iH\Delta t}$ with a single-timestep Trotterized unitary denoted $U = U(\Delta t)$.
\item  Variationally search for an approximate diagonalization of $U$ by compiling it to a unitary with a structure of the form
\begin{align}
\label{eqn:Vansatz}
    V(\alv , \Delta t) := W(\thv) D(\gav , \Delta t) W(\thv)^\dagger\,,
\end{align}
where $\alv = (\thv , \gav)$ is a vector of parameters. Here, $D(\gav , \Delta t)$ is a parameterized unitary that will (after training) encode the eigenvalues of $U(\Delta t)$, while $W(\thv)$ is a parameterized unitary matrix that will consist of the corresponding eigenvectors~\cite{cirstoiu2020variational}. The compilation is performed using the local Hilbert-Schmidt test~\cite{khatri2019quantum} to find the parameters $\thv_{\rm opt}$ and $\gav_{\rm opt}$ that minimize the local Hilbert-Schmidt cost. 

\item Use the compiled form to simulate for time $T = N\Delta t$ using the circuit 
\begin{align}
    W(\thv_{\rm opt}) D(\gav_{\rm opt} , N \Delta t) W(\thv_{\rm opt})^\dagger \, .
\end{align}
\end{enumerate}

VFF has proven effective for providing a fixed quantum circuit structure with which to fast-forward beyond the coherence time of current noisy quantum devices. However, the algorithm requires a full diagonalization of $U$ over the entire Hilbert space. The local Hilbert-Schmidt test used to find this diagonalization requires $2n$ qubits. Additionally, the ansatz must be sufficiently expressible to diagonalize the full unitary $U$ to a high degree of approximation~\cite{Sukin2019Expressibility,Nakaji2020Expressibility,holmes2021connecting}. This typically requires a large number of parameters and a reasonably deep circuit. These overheads limit VFF’s utility on current hardware.

\begin{figure*}[t!]
    \includegraphics[width=0.95\textwidth]{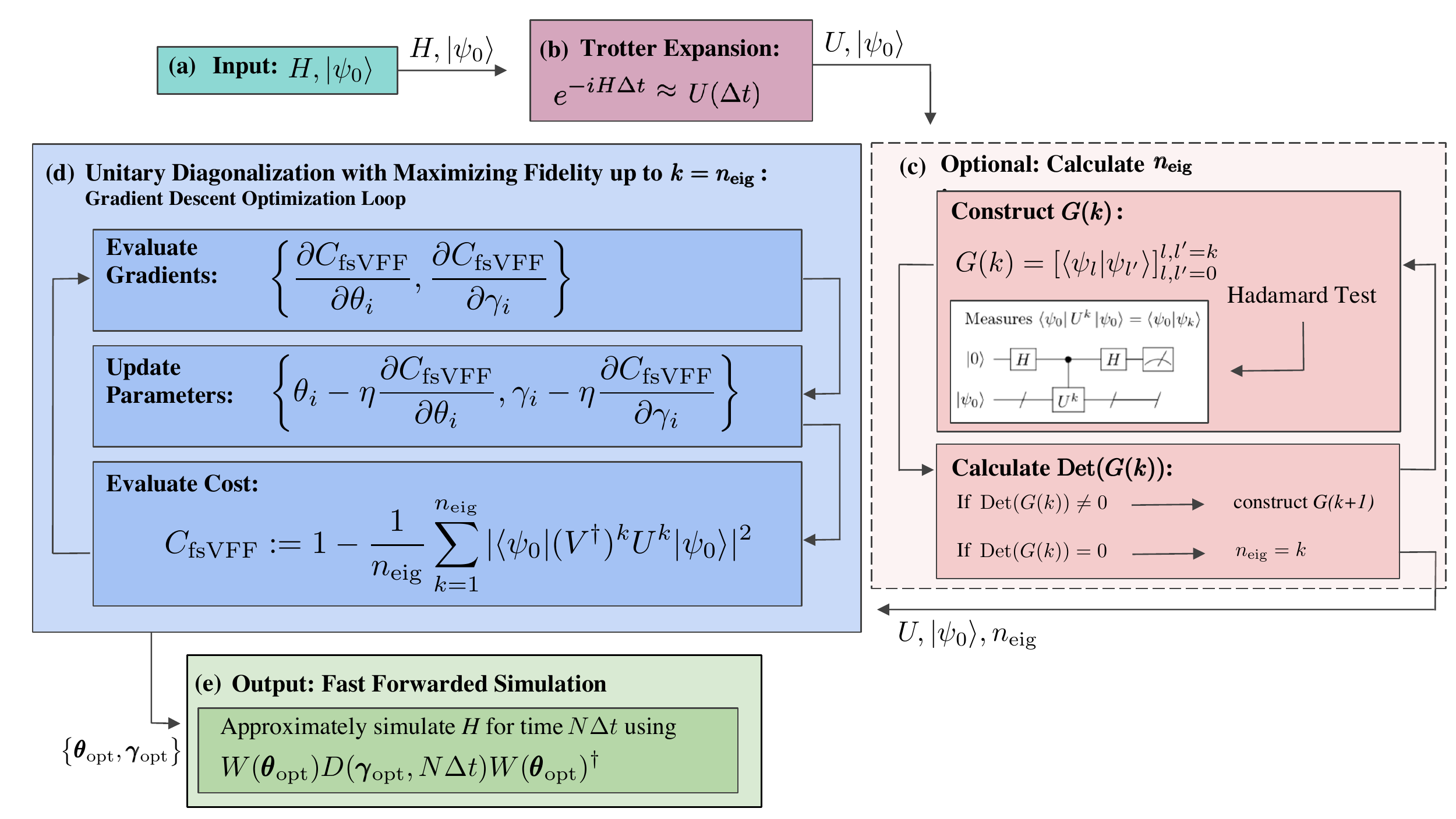}
    \caption{\textbf{The fsVFF Algorithm.}  (a) An input Hamiltonian and an initial input state are necessary (b) to create a single time-step Trotterized unitary, $ U(\Delta t)$  and (c) to calculate the number of eigenstates spanned by the initial state. The value of $n_{\rm eig}$ can be calculated by constructing a matrix of state overlaps $U^k \ket{\psi_0 }$ and increasing the matrix dimension until the determinant is zero. (d) The unitary is then variationally diagonalized into the form, $V(\alv, \Delta t) = W(\thv) D(\gamv, \Delta t) W^\dagger(\thv)$. The cost function $C_{\rm fsVFF}$ is minimized with a classical optimizer (e.g., gradient descent), where the parameters $\vec{\theta}$ and $\vec{\gamma}$ are updated. (e) The optimal parameters $\vec{\theta}_{\rm opt}$ and $\vec{\gamma}_{\rm opt}$ are then used to implement a fast-forwarded simulation with the diagonalized unitary form.} 
    \label{fig:flowchart}
\end{figure*}

In what follows, we introduce a more NISQ-friendly refinement to VFF that reduces these overheads when one is interested in fast-forwarding a \textit{fixed} initial state $\ket{\psi_0}$, rather than any possible initial state. The fixed state VFF algorithm (fsVFF) is summarised in Fig.~\ref{fig:flowchart}.

We note that VFF, like the standard iterated Trotter approach to quantum simulation, necessarily incurs a Trotter error by approximating $e^{-iH\Delta t}$ with $U = U(\Delta t)$. This Trotter error may be removed using the Variational Hamiltonian Diagonalization algorithm (VHD), which directly diagonalizes the Hamiltonian $H$~\cite{commeau2020variational}. However, VHD is yet more resource intensive than VFF on current hardware, so we focus here on refining VFF.


\section{Fixed State Variational Fast Forwarding Algorithm}\label{sec:VFF}

\subsection{Cost function}\label{sec:Cost}



In fsVFF, instead of searching for a full diagonalization of $U$ over the entire Hilbert space, we search for a diagonal compilation of $U$ that captures the action of $U$ on the initial state $\ket{\psi_0}$ and its future evolution, $e^{-i H t} \ket{\psi_0}$. Here, we introduce a cost function tailored to this task. 


To make precise what is required of the cost for fsVFF, let us first note that as the state $\ket{\psi_0}$ evolves, it remains within its initial energy subspace. This can be seen by expanding the initial state in terms of the energy eigenbasis $\{ \ket{E_k} \}_{k=1}^{2^n}$ (the eigenbasis of $H$) as 
\begin{equation}
    \ket{\psi_0} = \sum_{k=1}^{n_{\rm eig}} a_k \ket{E_k}, 
\end{equation}
where $a_k = \braket{E_k | \psi_0}$, and noting that
\begin{equation}
   e^{-i H t} \ket{\psi_0} = \sum_{k=1}^{n_{\rm eig}} a_k e^{-i E_k t} \ket{E_k} \, .
\end{equation}
Thus it follows that if $\ket{\psi_0}$ spans $n_{\rm eig}$ energy eigenstates of $H$, so does $e^{-i H t} \ket{\psi_0}$ for all future times. Therefore to find a compilation of $U$ that captures its action on $e^{-i H t} \ket{\psi_0}$ (for all times $t$) it suffices to find a compilation of $U$ on the $n_{\rm eig}$ dimensional subspace spanned by $\{ \ket{E_k} \}_{k =1}^{n_{\rm eig}}$. We stress that the eigenstates $\{ \ket{E_k} \}_{k=1}^{2^n}$ need not be ordered, and therefore the subspace spanned by the subset $\{ \ket{E_k} \}_{k =1}^{n_{\rm eig}}$ is not necessarily low lying in energy. 


A No-Free-Lunch Theorem for quantum machine learning introduced in Ref.~\cite{poland2020no} proves that to perfectly learn the action of a unitary on a $d$-dimensional space requires $d$ training pairs. In the context of fsVFF, we are interested in learning the action of a unitary on an $n_{\rm eig}$-dimensional subspace. Since the unitary is block diagonal, one can directly apply this NFL theorem to the subspace of interest. Therefore $n_{\rm eig}$ training pairs are required to learn the unitary's action on this subspace. (Note, we assume here that the training states are not entangled with an additional register. It was shown in Ref.~\cite{sharma2020reformulation} that using entangled training data can reduce the required number of training states. In fact, this more powerful method is used by the VFF algorithm. However, producing such entangled training data requires additional qubits and two-qubit gates and therefore is less NISQ-friendly.) 

The No-Free-Lunch theorem therefore implies that $n_{\rm eig}$ states are required to learn $U$ on $\ket{\psi_0}$ (assuming leakage due to Trotter error is negligible). In general these states may be freely chosen from the subspace spanned by $\ket{\psi_0}$. Here a convenient choice in training states would be $\ket{\psi_0}$ and its Trotter evolutions, that is the set $\{ U^k \ket{\psi_0} \}_{k=1}^{n_{\rm eig}}$. Motivated by these observations, we define our cost function for fsVFF as 
\begin{equation}\label{eq:fsVFFcost}
    C_{\rm fsVFF} := 1 - \frac{1}{n_{\rm eig}}\sum_{k=1}^{n_{\rm eig}} | \bra{\psi_0} (V^\dagger)^k U^k \ket{\psi_0} |^2 \, ,
\end{equation}
where similarly to VFF we use a diagonal ansatz $V(\alv , \Delta t) := W(\thv) D(\gav , \Delta t) W(\thv)^\dagger$.
This cost quantifies the overlap between the initial state evolved under $U$ for $k$ time steps, $U^k \ket{\psi_0}$, and the initial state evolved under the trained unitary for $k$ time steps, $W D^k W^\dagger \ket{\psi_0}$, averaged over $n_{\rm eig}$ time steps. Assuming we have access to the unitary that prepares the state $\ket{\psi_0}$, the state overlaps can be measured using $n$ qubits, via a circuit that performs a Loschmidt echo~\cite{sharma2019noise}. Therefore $C_{\rm fsVFF}$ can be evaluated using only $n$ qubits. This is half as many as standard VFF, opening up the possibility of performing larger simulations on current hardware. 

It is important to note that while the exact time-evolved state $\exp(-i H t) \ket{\psi_0}$ is perfectly confined to the initial subspace, the approximate evolution induced by $U(\Delta t)$ allows for leakage from the initial subspace~\cite{sahinoglu2020hamiltonian}. Thus the subspace spanned by $\{U^k \ket{\psi_0}\}_{k =1}^{n_{\rm eig}}$ in general does not perfectly overlap with $\{ \ket{E_k} \}_{k =1}^{n_{\rm eig}}$. However, by reducing $\Delta t$ and considering higher order Trotter approximations \cite{sornborger1999higher, suzuki1976generalized}, this leakage can be made arbitrarily small. In Appendix~\ref{ap:Faithful}, we prove that in the limit that leakage from the initial subspace is negligible, $C_{\rm fsVFF}$ is faithful. That is, we show that the cost vanishes, $C_{\rm fsVFF} = 0$, if and only if the fidelity of the fast-forwarded simulation is perfect,
\begin{equation}
  F_\tau =  | \bra{\psi_0}  W^\dagger D^\tau W U^\tau \ket{\psi_0} |^2 = 1 \, ,
\end{equation}
for all times $\tau$. Note, that the reverse direction is trivial. If $F_\tau =1$ for all $\tau$, then $C_{\rm fsVFF} = 0$.

Similar to the VFF cost, the fsVFF cost is noise resilient in the sense that incoherent noise should not affect the global optimum of the function. This is proven for a broad class of incoherent noise models using the results of Ref.~\cite{sharma2019noise} in Appendix~\ref{ap:NoiseResilient}. 

Nonetheless, it is only possible to measure $C_{\rm fsVFF}$ if the unitary $U^{n_{\rm eig}}$ can be implemented comfortably within the coherence time of the QC. 
Additionally, the number of circuits required to evaluate $C_{\rm fsVFF}$ in general scales with $n_{\rm eig}$. Given these two restrictions, fsVFF is limited to simulating quantum states spanning a non-exponential number of eigenstates. Consequently, we advocate using fsVFF to simulate states with $n_{\rm eig} = \text{poly}(n)$. Crucially these states need {\it not} be low lying and therefore our approach is more widely applicable than the Subspace Variational Quantum Simulator (SVQS) algorithm~\cite{heya2019subspace}, which simulates fixed low energy input states. In Section~\ref{ap:LargeNeig} we develop methods for reducing the resources required to evaluate the cost for larger values of $n_{\rm eig}$.


While $C_{\rm fsVFF}$ was motivated as a natural choice of cost function to learn the evolution induced by a target unitary on a fixed initial state, it is a global cost~\cite{cerezo2020cost} and hence it encounters what is known as a \textit{barren plateau} for large simulation sizes~\cite{mcclean2018barren,cerezo2020cost,cerezo2020impact,arrasmith2020effect,Holmes2020Barren,holmes2021connecting,volkoff2021large,sharma2020trainability,pesah2020absence,uvarov2020barren,marrero2020entanglement,patti2020entanglement}. In Appendix~\ref{ap:LocalCost} we suggest an alternative local version of the cost to mitigate such trainability issues.

\subsection{Calculating $n_{\rm eig}$}\label{sec:Neig}


In this section, we present an algorithm to calculate $n_{\rm eig}$ and therefore determine the number of training states required to evaluate $C_{\rm fsVFF}$. 
Our proposed algorithm utilizes the fact that the number of energy eigenstates spanned by $\ket{\psi_0}$ is equivalent to the number of linearly independent states in the set $\mathcal{V}_\infty$ where $\mathcal{V}_k := \{ \ket{\psi_l} \}_{l=0}^{l=k} $ with $ \ket{\psi_l} := U(\Delta t)^l \ket{\psi_0}$. The subspace $\mathcal{K}_k(U, \psi_0)$ spanned by $\mathcal{V}_k$ is known as the Krylov subspace associated with the operator $U$ and vector $\ket{\psi_0}$~\cite{krylov1931numerical}. Therefore, $n_{\rm eig}$ is equivalently the dimension of the Krylov subspace $\mathcal{K}_\infty(U, \psi_0)$.

To determine the dimension of $\mathcal{K}_\infty(U, \psi_0)$ we can utilize the fact that the determinant of the Gramian matrix of a set of vectors (i.e., the matrix of their overlaps) is zero if and only if the vectors are linearly dependent.
The Gramian corresponding to $\mathcal{V}_k$ is given by
\begin{equation}
    G(k) = 
    \begin{pmatrix}
 \braket{\psi_0|\psi_0} & \braket{\psi_0|\psi_1}  & \cdots & \braket{\psi_0|\psi_k}  \\
\braket{\psi_1|\psi_0} & \braket{\psi_1|\psi_1} & \cdots & \braket{\psi_1|\psi_k}\\
\vdots & \vdots & \ddots & \vdots \\
\braket{\psi_k|\psi_0} & \braket{\psi_k|\psi_1} & \cdots & \braket{\psi_k|\psi_k}
\end{pmatrix} \, .
\end{equation}
If $\text{Det}(G(k)) \neq 0$, then the vectors in $\mathcal{V}_k$ are linearly independent and therefore span \textit{at least} a $k+1$ dimensional subspace. Conversely, if $\text{Det}(G(k)) = 0$, the set $\mathcal{V}_k$ contains linear dependencies and the subspace they span is less than $k+1$ dimensional. Therefore, if we can find $k_{\rm min}$, the smallest $k$ such that $\text{Det}(G(k)) = 0$, then (noting that $G(k)$ is a $k+1$ dimensional matrix) we know that $k_{\rm min}$ is the largest number of linearly independent vectors spanned by $\mathcal{V}_{\infty}$. That is, $k_{\rm min}$ is the dimension of $\mathcal{K}_\infty(U, \psi_0)$ and so we have that $n_{\rm eig} = k_{\rm min}$.

The overlaps $\braket{\psi_l|\psi_{l'}}$ for any $l$ and $l'$ can be measured using the Hadamard Test, shown in Fig.~\ref{fig:flowchart}, and thus the Hadamard test can be used to determine $G(k)$ on quantum hardware. Since the Gramian here contains two symmetries, hermiticity and the invariance $\braket{\psi_l|\psi_{l'} } = \langle \psi_0| U^{-l} U^{l'} |\psi_0 \rangle = \langle \psi_0 | U^{{l'}-l} |\psi_0 \rangle = \langle \psi_0 | \psi_{{l'}-l} \rangle$, we only have to calculate the first row of the matrix $G(k)$ on the quantum computer. 

In summary, our proposed algorithm to determine $n_{\rm eig}$ consists of the following loop. Starting with $k = 1$,
\begin{enumerate}
    \item Construct $G(k)$ using the Hadamard test. 
    \item Calculate (classically) $\text{Det}(G(k))$.
    
    If $\text{Det}(G(k)) = 0$, terminate the loop and conclude that $n_{\rm eig} = k$.
    
    If $\text{Det}(G(k)) \neq 0$, increase $k \rightarrow k +1$ and return to step 1.

\end{enumerate}
This is shown schematically in Fig.~\ref{fig:flowchart}. 

\medskip

We remark that in the presence of degeneracies in the spectrum of $H$, the eigenvectors corresponding to degenerate eigenvalues are not unique. Therefore, in this case, the number of states spanned by $\ket{\psi_0}$ depends on how the eigenvectors corresponding to degenerate eigenvalues are chosen. However, as detailed in Appendix~\ref{ap:Faithful}, to learn the action of $U$ on $\ket{\psi_0}$, what matters is the number of eigenstates spanned by $\ket{\psi_0}$ corresponding to \textit{unique} eigenvalues. This is equivalent to the dimension of the Krylov subspace $\mathcal{K}_\infty(U, \psi_0)$. Consequently, the algorithm detailed above can also be used in this case. 

\medskip

While it is beneficial to learn $n_{\rm eig}$ to determine how many training states are required to perfectly learn the diagonalization on the subspace spanned by the initial state, we stress that it is not strictly necessary for the successful implementation of fsVFF. One could always train on an increasing number of states and study the convergence of an observable of interest. More concretely, one could train on $k$ states and then use the resultant diagonalization to compute the evolution of a particular observable as a function of time. For $k < n_{\rm eig} $ the trajectory of the observable will alter as $k$ is increased. However, for $k \geq n_{\rm eig}$ increasing $k$ further will no longer change the trajectory of the observable because it will have already converged on the true trajectory. Using this approach, $n_{\rm eig}$ need not be already known to implement fsVFF.





\subsection{Ansatz}\label{sec:Ansatz}

The fsVFF algorithm, similarly to VFF, employs an ansatz of the form 
\begin{align}\label{eq:VHDansatz}
V(\vec{\alpha}, \Delta t)= W(\thv)D(\gamv, \Delta t)W^\dagger(\thv) \, ,
\end{align}
to diagonalize the initial Trotter unitary $U(\Delta t)$. Here $W(\thv)$ is a quantum circuit that approximately rotates the standard basis into the eigenbasis of $H$, and $D(\gamv)$ is a diagonal unitary that captures the (exponentiated) eigenvalues of $H$. A generic diagonal operator $D$ can be written in the form
\begin{equation}\label{eq:D}
    D(\gamv, \Delta t) = \prod_{\vec{q}} e^{i \gamma_{\vec{q}} Z^{\vec{q}} \Delta t} \, ,
\end{equation}
where $\gamma_{\vec{q}}\in \mathbb{R}$ and we use the notation
\begin{equation} \label{eq:Zq}
Z^{\vec{q}}=Z_1^{q_1}\otimes\cdots\otimes Z_n^{q_n}\,,
\end{equation}
with $Z_j$ the Pauli $Z$ operator acting on qubit $j$. 
While Eq.~\eqref{eq:D} provides a general expression for a diagonal unitary, for practical ans\"{a}tze it may be desirable to assume that the $Z^{\vec{q}}$ operators are local operators and the product contains a polynomial number of terms, i.e., is in $\mathcal{O}(\text{poly}(n))$. There is more flexibility in the construction of the ans\"{a}tze for $W$ since these are generic unitary operations. A natural choice might be to use a hardware-efficient ansatz~\cite{kandala2017hardware} or an adaptive ansatz~\cite{khatri2019quantum, bilkis2021semi}. 

One of the main advantages of fsVFF is that diagonalization is only necessary over the subspace spanned by the initial state, rather than the entire Hilbert space which will be significantly larger. To outperform standard VFF, it is in our interest to take advantage of this small subspace to find compact ans\"{a}tze.

The two main impeding factors we wish to minimize to aid diagonalization are error rates and optimization time. Therefore, when searching for ans\"{a}tze, our priorities are to minimize the number of CNOT gates required (the noisiest component in the ans\"{a}tze) and the number of rotation parameters. There is, however, a trade off between expressibility of the ansatz and its trainability. There needs to be enough freedom in the unitary to map the required eigenvectors to the computational basis but generically highly expressive ans\"{a}tze exhibit barren plateaus~\cite{holmes2021connecting}. 

For systems with symmetries and/or systems that are nearby perturbations of known diagonalizable systems, it may be possible to find a fully expressive, compact ansatz by inspection. This is the case for a simple 2-qubit XY Hamiltonian, as discussed in Section~\ref{sec:Implementations}. 

More generally, it can be challenging to analytically find compact but sufficiently expressible ans\"{a}tze. Nonetheless, it is possible to variationally update the ansatz structure and thereby systematically discover simple structures. One straightforward approach is to use a layered ansatz where each layer initializes to the identity gate~\cite{grant2019initialization,skolik2020layerwise}. The ansatz can be optimized until it plateaus, redundant single qubit gates removed, then another layer can be appended and the process repeats. Alternatively, more sophisticated discrete optimization techniques may be used to variationally search the space of ans\"{a}tze.

\subsection{Summary of algorithm}

The fixed state Variational Fast Forwarding algorithm (fsVFF) is summarized in Fig.~\ref{fig:flowchart}. We start with an initial state $\ket{\psi_0}$ that we wish to evolve under the Hamiltonian~$H$. 

\begin{enumerate}
    \item The first step is to approximate the short time evolution using a single step Trotter approximation $U$.
    \item This Trotter approximation can be used to find an approximation for $n_{\rm eig}$, the dimension of the energy eigenspace spanned by $\ket{\psi_0}$, using the method outlined in Section~\ref{sec:Neig}.
    \item Equipped with a value for $n_{\rm eig}$, we then variationally search for a diagonalization of $U$ over the energy subspace spanned by $\ket{\psi_0}$ using $C_{\rm fsVFF}$, Eq.~\eqref{eq:fsVFFcost}.
    At each iteration step the gradient of the cost with respect to a parameter $\theta_i$ is measured on the quantum computer for a fixed set of parameters using the analytic expressions for $\partial_{\theta_i} C_{\rm fsVFF}$ provided in Appendix~\ref{ap:Gradients}. These gradients are used to update the parameters using a classical optimizer, such as those in Refs.~\cite{kubler2020adaptive,arrasmith2020operator,sweke2020stochastic}. The output of the optimization loop is the set of parameters that minimize $C_{\rm fsVFF}$, 
    \begin{align}\label{eq:optimization}
    \{\thv_{\opt},\gamv_{\opt}\}=\underset{\thv,\gamv}{\text{ arg min }} C_{\rm fsVFF}(\thv,\gamv) \, .
    \end{align}
\item Finally, the state $\ket{\psi_0}$ can be simulated for time $T = N \Delta t$ using the circuit
\begin{align}
    W(\thv_{\rm opt}) D(\gav_{\rm opt} , N \Delta t) W(\thv_{\rm opt})^\dagger \, .
\end{align}
That is, by simply multiplying the parameters $\gav_{\opt}$ in the diagonalized unitary by a constant number of iterations $N$.
\end{enumerate}

In Appendix~\ref{ap:errors}, we show that the total simulation fidelity, in the limit that leakage is small, is expected to scale sub-quadratically with the number of fast-forwarding time steps $N$. Thus, if the minimal cost from the optimization loop is sufficiently small, we expect the fsVFF algorithm to allow for long, high fidelity simulations.

\section{Hardware Implementation}\label{sec:Implementations}





In this section we demonstrate that fsVFF can be used to implement long time simulations on quantum hardware. Specifically, we simulate the XY spin chain, which has the Hamiltonian
\begin{equation}
    H_{\rm XY} := \sum_{j=1}^{n-1} X_{j} X_{j+1} + Y_{j} Y_{j+1} \, ,
\end{equation}
where $X_j$ and $Y_j$ are Pauli operators on the $j_{\rm th}$ qubit. 
In what follows, we first present results showing that we can determine $n_{\rm eig}$ for an initial state $\ket{\psi_0}$ using the method described in Section~\ref{sec:Neig}. We then demonstrate that the fsVFF cost can be trained to find an approximate diagonalization of $H_{\rm XY}$ on the subspace spanned by $\ket{\psi_0}$. We finally use this diagonalization to perform a long time fast forwarded simulation. In all cases we focus on a two qubit chain, i.e. $n=2$, and we approximate its evolution operator using a first-order Trotter approximation.

\subsection{Determining $n_{\rm eig}$}

\begin{figure}
\centering
\includegraphics[width =\columnwidth]{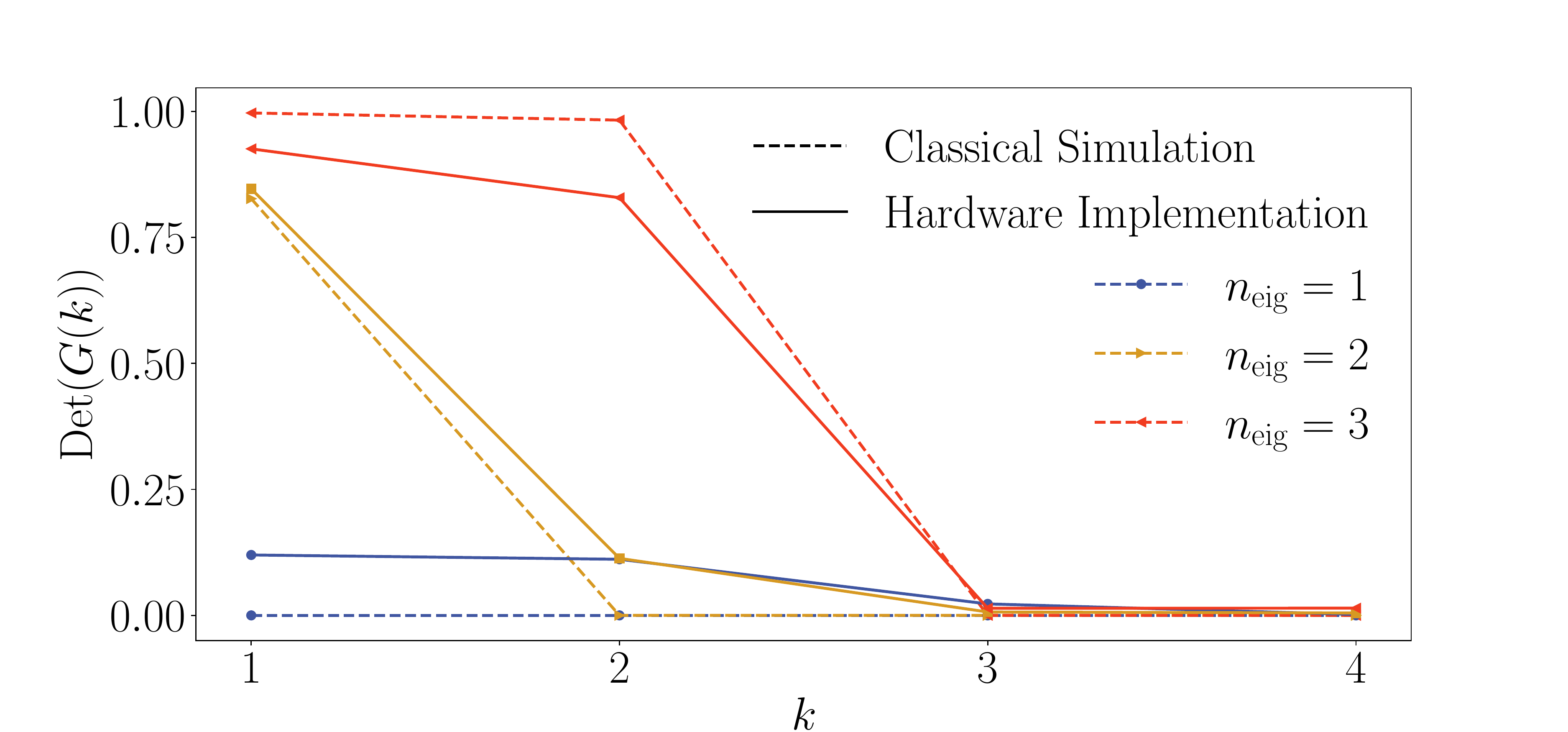}
\vspace{-2mm}
\caption{\textbf{Gramian Determinant Calculation.} Here we plot the determinant of the Gramian matrix, $\text{Det}(G)$, for $G$ measured on the Honeywell quantum computer (solid) and simulated classically (dashed) for a 2-qubit XY spin chain. Specifically we looked at states spanning $k = 1$ (blue),  $k=2$ (yellow) and $k = 3$ (red) eigenstates. For both sets of data $\text{Det}(G(n_{\rm eig} )) \approx 0$, demonstrating the effectiveness of the method for determining $n_{\rm eig}$ that we introduce in Section~\ref{sec:Neig}. For the Honeywell implementation we used 1000 measurement samples per circuit.}
\label{fig:GramianDeterminant}
\end{figure}

The 2-qubit XY Hamiltonian has the eigenvectors $\{\ket{00},  \frac{1}{\sqrt{2}}(\ket{10}+\ket{01}),  \frac{1}{\sqrt{2}}(\ket{10}-\ket{01}), \ket{11}\}$, corresponding to the eigenvalues \{0, 1, -1, 0\}. As proof of principle, we tested the algorithm for determining $n_{\rm eig}$ on the states $\ket{00}$ (corresponding to $n_{\rm eig} = 1$), $\ket{10}$ ($n_{\rm eig}=2$) and $\frac{1}{\sqrt{2}}(\ket{00}+ \ket{10})$ ($n_{\rm eig} = 3$). As described in Section~\ref{sec:Neig}, the $n_{\rm eig}$ of these states can be found by calculating $\text{Det}(G(k))$ for increasing values of $k$ since, as $k$ is increased, the determinant first equals 0 when $k = n_{eig}$.

To verify this for the states considered here, we first determine $G$ using a classical simulator. As seen in Figure~\ref{fig:GramianDeterminant}, in this case $\text{Det}(G(k))$ exactly equals 0 when $k = n_{eig}$. We then measured $G$ on Honeywell's quantum computer. Although on the real quantum device gate noise and sampling errors are introduced, the results reproduce the classical results reasonably well. Namely, at the correct value of $k$, $\text{Det}(G(k))$ drastically reduces and approximately equals 0. Thus, we have shown that it is possible to determine $n_{\rm eig}$ for an initial state by measuring $G$ on quantum hardware.  

\subsection{Training}
We tested the training step of the algorithm on IBM and Rigetti's quantum computers, specifically ibmq\_toronto and Aspen-8.
For the purpose of implementing a complete simulation, we chose to focus on simulating the evolution of the state $|\psi_0\rangle = |10\rangle$. As discussed in the previous section, this state spans $n_{\rm eig} = 2$ eigenstates.

To diagonalize $H_{\rm XY}$ on the 2-dimensional subspace spanned by $|10\rangle$, we used a hybrid quantum-classical optimization loop to minimize $C_{\rm fsVFF}$. For a state with $n_{\rm eig} = 2$ the cost $C_{\rm fsVFF}$, Eq.~\eqref{eq:fsVFFcost}, uses two training states $\{U(\Delta t)^k|\psi_0\rangle\}_{k=1,2}$ where $U(t)$ is the first-order Trotter approximation of $H_{\rm XY}$.
On the IBM quantum computer we evaluated the full cost function for each gradient descent iteration. However, the time available on the Aspen-8 device was limited, so to speed up the rate of optimization we evaluated the overlap on just one of the two training states per iteration, alternating between iterations (instead of evaluating the overlaps on both training states every iteration). To allow the movement through parameter space to use information averaged over the two timesteps, whilst only using a single training state per cost function evaluation, momentum was added to the gradient updates \cite{defazio2020understanding}. 

\begin{figure}[t]
\centering
\includegraphics[width =\columnwidth]{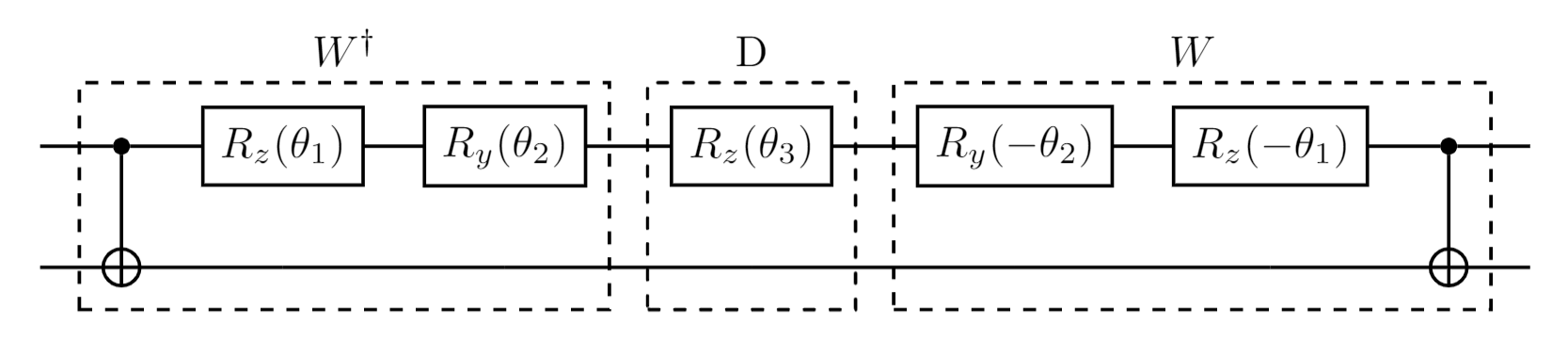}
\vspace{-2mm}
\caption{\textbf{Ansatz for Hardware Implementation}. The ansatz used to diagonalize the 2-qubit XY Hamiltonian in the subspace of initial state $\ket{10}$ for the implementation on Rigetti and IBM's quantum computers. Here $R_j(\theta) = \exp(- i \theta \sigma_j/2)$ for $j = x, y, z$.}
\label{fig:AnsatzTwoQubit}
\end{figure}

To take advantage of the fact that more compact ans\"{a}tze are viable for fsVFF, we variationally searched for a short depth ansatz, tailored to the target problem. Specifically, we started training with a general 2-qubit unitary and then during training the structure was minimised by pruning unnecessary gates. In Figure~\ref{fig:AnsatzTwoQubit}, we show the circuit for the optimal ansatz obtained using the method. The ansatz requires one CNOT gate and two single qubit gates for $W$ and only one $R_z$ rotation for $D$. This is a substantial compression on the most general two qubit ansatz for $W$ which requires 3 CNOTs and 15 single qubit rotations and the most general 2 qubit ansatz for $D$ which requires 2 $R_z$ rotations and one 2-qubit ZZ rotation (though in the case of the XY Hamiltonian this may be simplified to only 2 $R_z$ rotations \cite{lieb1961two}).

\begin{figure}[t!]
\centering
\includegraphics[width =\columnwidth]{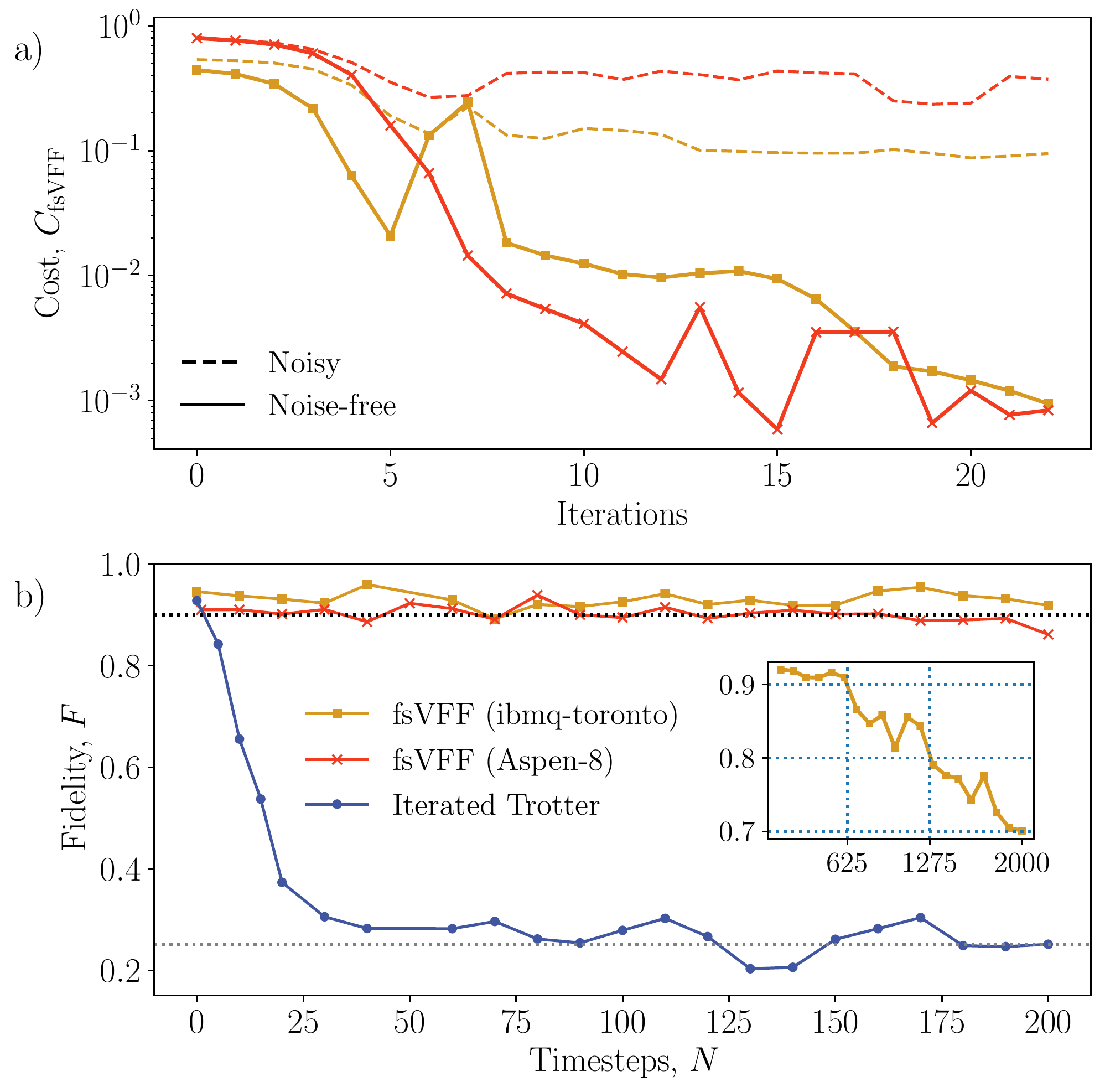}
\vspace{-2mm}
\caption{\small \textbf{Hardware Implementation.} a) The 2-qubit parameterised quantum circuit shown in Figure~\ref{fig:AnsatzTwoQubit} was trained to diagonalize $U(\Delta t)$, a first order Trotter expansion of the 2-qubit XY Hamiltonian with $\Delta t = 0.5$, in the subspace spanned by $|10\rangle$. The dashed line plots the noisy cost as measured on ibmq\_toronto (yellow) and Aspen-8 (red) using 30,000 samples per circuit. The solid line indicates the equivalent noise-free cost that was calculated on a classical simulator. b) The initial state $|\psi_0\rangle = |10\rangle$ is evolved forwards in time on the ibmq\_rome quantum computer using the iterated Trotter method (blue) and using fsVFF with the optimum parameters found on ibmq\_toronto (yellow) and Aspen-8 (red). The quality of the simulation is evaluated by plotting the fidelity $F = \langle \psi | \rho | \psi \rangle$ between the evolved state and exact evolution. The grey dotted line at $F = 0.25$ represents the overlap with the maximally mixed state. The black dotted line denotes a threshold fidelity at $F = 0.9$. The inset shows the fast-forwarding of the ansatz trained on ibmq\_toronto on a longer timescale, where the fidelity dropped below 0.9 (0.8) at 625 (1275) timesteps. All simulation data was taken using 1000 samples per circuit.}
\label{fig:HardwareFastForwarding}
\end{figure}

Figure~\ref{fig:HardwareFastForwarding}(a) shows the fsVFF cost function versus the number of iterations for the implementations on ibmq\_toronto (yellow) and Aspen-8 (red). The dashed line indicates the noisy cost value obtained from the quantum computer. To evaluate the quality of the optimization, we additionally classically compute the true cost (indicated by the solid lines) using the parameters found on ibmq\_toronto and Aspen-8.
While the noisy cost saturates at around $10^{-1}$, we obtained a minimum noise-free cost of the order $10^{-3}$.
The two orders of magnitude difference between the noisy and the noise-free cost is experimental evidence that the cost function is noise resilient on extant quantum hardware. 

\subsection{Fast forwarding}

Finally we took the two sets of parameters found from training on ibmq\_toronto and Aspen-8, and used them to implement a fast-forwarded simulation of the state $\ket{10}$ on ibmq\_rome. 
To evaluate the quality of the fast forwarding we calculated the fidelity, $F(N) = \langle\psi(N)|\rho(N)|\psi(N)\rangle$, between the density matrix of the simulated state, $\rho(N)$, after $N$ timesteps, and the exact time evolved state, $|\psi(N)\rangle$, at time $T = N \Delta t$. We used Quantum State Tomography to reconstruct the output density matrix, and then calculated the overlap with the exact state classically.


As shown by the plots of $F(N)$ in Figure~\ref{fig:HardwareFastForwarding}(b), fsVFF significantly outperforms the iterated Trotter method. 
Let us refer to the time before the simulation falls below an error threshold $\delta$ as the high fidelity time. Then the ratio of the high fidelity time for fsVFF ($T^{\rm FF}_\delta$) and for standard Trotterization ($T^{\rm Trot}_\delta$) is a convenient measure of simulation performance,
\begin{equation}
    R^{\rm FF}_\delta = T^{\rm FF}_\delta / T^{\rm Trot}_\delta \, .
\end{equation}
A simulation can be said to have been successfully \textit{fast-forwarded} if $R^{\rm FF}_\delta > 1$. The iterated Trotter method dropped below a simulation infidelity threshold of $\delta = 1-F = 0.1 \ (0.2)$ after 4 (8) timesteps. In comparison, fsVFF maintained a high fidelity for 625 (1275) timesteps. Thus we achieved a simulation fast-forwarding ratio of $R_{0.1}^{\rm FF} = 156$ ($R_{0.2}^{\rm FF} = 159$).

\section{Numerical Simulations}
\subsection{Noisy Training}

We further validate fsVFF's performance by testing it on a simulator of a noisy quantum computer. The noise levels on the simulator are lower than those experienced on current devices and hence these results are indicative of the performance of the algorithm in the near future as hardware improves. 

For these numerics we diagonalize the evolution of the 4 qubit XY Hamiltonian, in the subspace spanned by the domain wall state $\ket{\psi_0} = \ket{1100}$. This state spans 5 energy eigenstates of the XY Hamiltonian and so we use the training states $\{U(\Delta t)^k \ket{\psi_0}\}_{k=1}^5$. Here $U(\Delta t)$ is chosen to be a second-order Trotter-Suzuki decomposition for the evolution operator under $H_{XY}$ with $\Delta t = 0.5$. The noise model used was based upon the IBM architecture.

To construct the ansatz for the diagonalizing  unitary, $W$, we developed an adaptive technique, similar to that proposed in ~\cite{khatri2019quantum, bilkis2021semi}, to evolve the discrete circuit structure, as well as optimize the rotation parameters using gradient descent. This method tends to produce shallower circuits than the ones obtained with fixed ansatz approaches. It is also less prone to get stuck in local minima.
Since, the XY Hamiltonian is particle number conserving we further use only particle number conserving gates. 
This reduces the number of parameters in $W$, as well as minimizing the leakage out of the symmetry sector when the circuit is executed with a noisy simulator.
Additional details on this adaptive learning method are provided Appendix~\ref{ap:Numerics}. The ansatz for $D$, as in our 2 qubit hardware implementation, simply consisted of $R_z$ rotations on each qubit. 

\begin{figure}[t]
\centering
\includegraphics[width =\columnwidth]{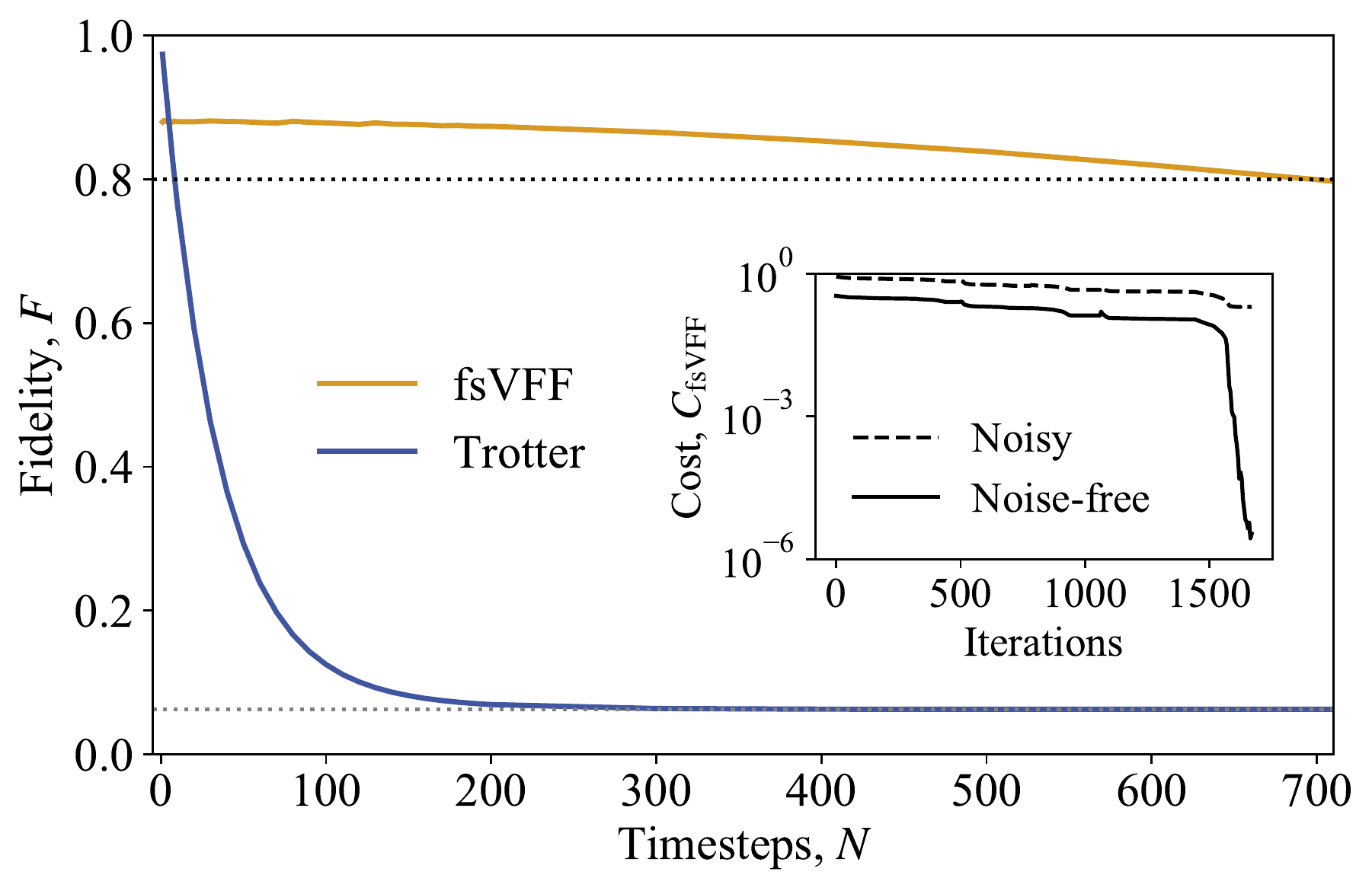}
\vspace{-5mm}
\caption{\textbf{Noisy Training and Fast-Forwarding of the 4 qubit XY Hamiltonian.} The inset shows the cost curve as the ansatz is evolved and optimized to diagonalize the 4 qubit XY Hamiltonian in the 5-dimensional subspace spanned by initial state $\ket{1100}$. The final circuit found by the learning algorithm for the diagonalizing unitary, $W$, had 50 CNOT gates. The main plot evaluates the fast-forwarding performance of the trained ansatz, in comparison to the Iterated-Trotter evolved state. The fidelity is calculated against the ideal state found in simulation using the iterated Trotter method in the absence of noise, $F(N) = \bra{\psi(N)}\rho\ket{\psi(N)}$ with $\ket{\psi(N)} = U(\Delta t)^N\ket{\psi_0}$. The black dotted line highlights a threshold value $F = 0.8$. The gray dotted line at $F = 1/2^4$ represents the overlap with the maximally mixed state.}
\label{fig:NoisyFF}
\end{figure}

The result of the training is shown in the inset of Fig.~\ref{fig:NoisyFF}. The noisy cost was measured by the noisy quantum simulator, whereas the noise-free cost is calculated simultaneously but in the absence of any noise. The significant separation between the noisy and noise-free cost again demonstrates the noise resilience of the VFF algorithm. 
After successfully training the cost, the fast-forward performance was then evaluated. Using the same noise model, the output density matrix of the Iterated-Trotter state and the fast-forwarded state was compared against the the Iterated-Trotter state in the absence of noise, with the fidelity between the two states plotted.
As shown in Fig.~\ref{fig:NoisyFF}, the fast-forwarded evolution significantly out performs the Iterated-Trotter evolution, with the former’s fidelity dropping below 0.8 after 700 timesteps, compared to only 8 steps of the latter. Thus we achieved a fast forwarding ratio of $R^{\text{FF}}_{0.8} = 87.5$.

\subsection{Fermi-Hubbard model}

Finally, to probe the scalability and the breadth of applicability of the fsVFF algorithm we performed a larger (noiseless) numerical implementation of the algorithm on the Fermi-Hubbard model. Specifically, we considered the 1D Fermi-Hubbard Hamiltonian on an $L$-site lattice with open boundary conditions:
\begin{equation} \label{eq:FH_Hamiltonian}
\begin{split}
H_\mathrm{FH}  = &  -J \sum_{j=1}^{L-1} \sum_{\sigma = \uparrow, \downarrow} 
c_{j,\sigma}^\dagger c_{j+1,\sigma} \ + \ \mathrm{h.c.} \\
 & + U \sum_{j=1}^L n_{j,\uparrow} n_{j,\downarrow} \ .
\end{split}
\end{equation}
Here, $c_{j,\sigma}$ ($c^\dagger_{j,\sigma}$) denotes fermionic creation (annihilation) operator at site $j$ for each of the two spin states $\sigma = \uparrow, \downarrow$ and $n_{j,\sigma} = c^\dagger_{j,\sigma} c_{j,\sigma}$ is a particle number operator. The total number of fermions with a spin $\sigma$ is given by $N_\sigma = \sum_j n_{j,\sigma}$. The term with coefficient $J$ in Eq.~\eqref{eq:FH_Hamiltonian} represents a single-fermion nearest-neighbor hopping and the term with coefficient $U$ introduces on-site repulsion. The Hamiltonian preserves particle numbers $N_\uparrow$ and $N_\downarrow$.

In our numerical studies, we choose $L=4$ (which requires 8 qubits to simulate) and $J=1$, $U=2$ as well as $N_\uparrow = N_\downarrow = 2$ (half filling). The initial state is chosen to be a superposition of $n_\mathrm{eig} = 5$ eigenvectors of $H_\mathrm{FH}$ in the particle sector $N_\uparrow = N_\downarrow = 2$. Similar to our noisy simulations of the XY model, we utilize an adaptive ansatz for $W$ that is made out of gates that preserve particle number $N_\uparrow$ and $N_\downarrow$. The ansatz for $D$ takes the form of Eq.~\eqref{eq:D}, where we only allow for single-$Z$ terms in Eq.~\eqref{eq:Zq}.
For this numerical result, we trained using the full exponentiation of the Hamiltonian as the evolution operator, with no Trotter error.

\begin{figure}[t!]
\centering
\includegraphics[width =\columnwidth]{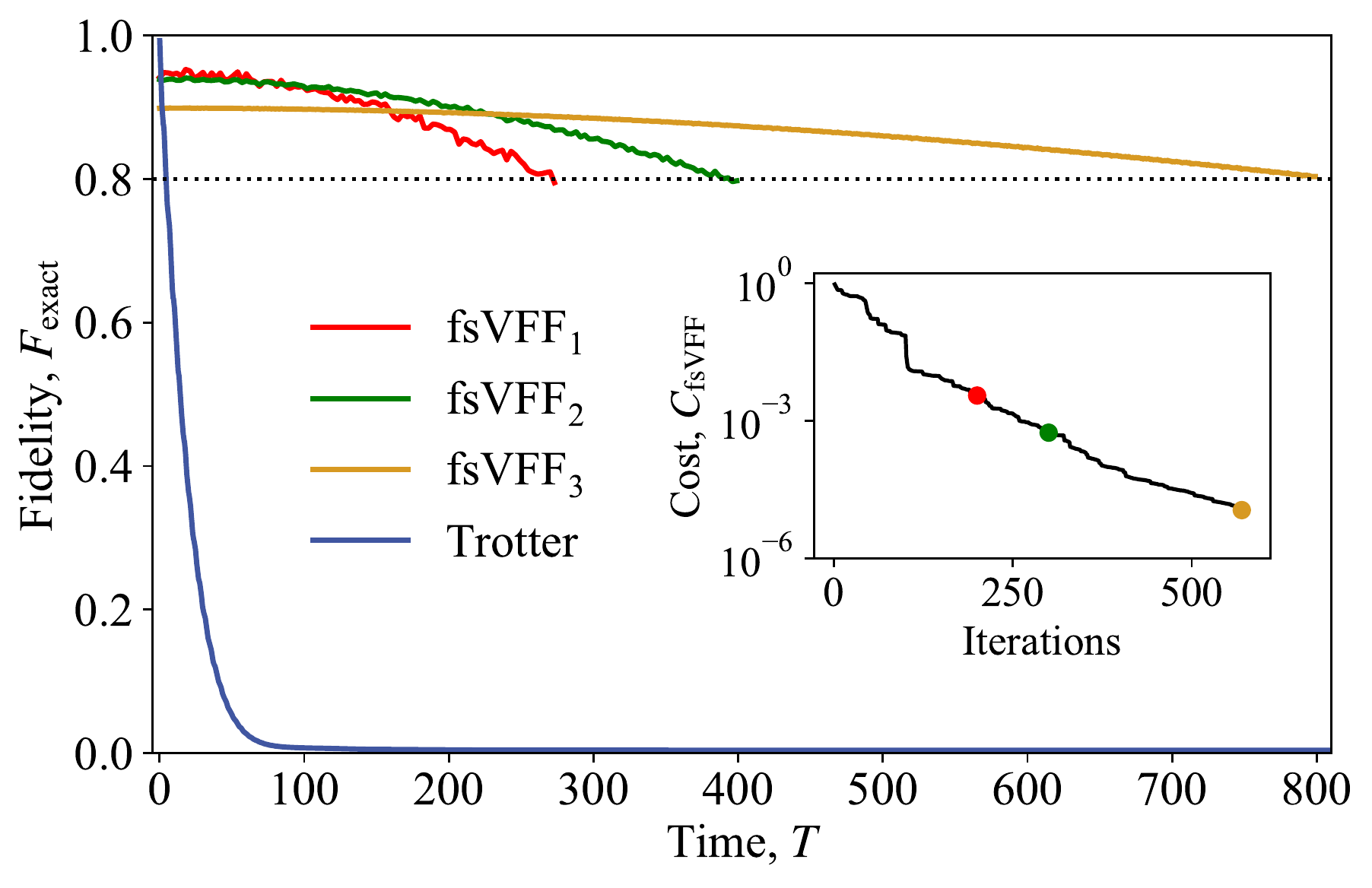}
\vspace{-2mm}
\caption{\textbf{Training and Fast-Forwarding of the 8 qubit Hubbard Model.} The inset shows the cost as it is iteratively minimized using an adaptive ansatz. Various quality diagonalizations are indicated by the colored circles. In the main figure, we plot the fidelity between the simulated state and the exact evolution as a function of time. The red, green and yellow lines denote fsVFF simulations using the corresponding quality diagonalization shown in the inset. In blue we plot the fidelity of the Iterated-Trotter simulation, $F_{\text{exact}}(T) = \bra{\psi(T)}\rho_{\rm trot}\ket{\psi(T)}$ with $\ket{\psi(T)} = e^{-iHT}\ket{\psi_0}$ and $\rho_{\rm trot}$ the simulated iterated Trotter state.}
\label{fig:Hubbard_FF}
\end{figure}


In the inset of Fig.~\ref{fig:Hubbard_FF} we show the cost function as it is iteratively minimized. We then test the performance on a noisy simulator based upon a fully connected 8-qubit trapped-ion device~\cite{trout2018simulating}. As shown in Fig.~\ref{fig:Hubbard_FF}, small final cost values typically require deeper circuits to achieve, the optimum diagonalization to use depends on the length of time one wishes to simulate. At short times, a larger final cost function value performs better since this corresponds to a shorter ansatz which experiences less noise. However, to simulate longer times, a higher quality diagonalization is required, with the additional noise induced by increased circuit depth resulting in a relatively small decrease in fidelity. As shown in Fig.~\ref{fig:Hubbard_FF}, we find that the fast forwarding corresponding to an optimized cost of $1.1 \times 10^{-5}$ maintained a fidelity of greater than 0.8 for $T < 800$. In contrast, the iterated Trotter method drops below 0.8 for $T > 4.6$ and hence we here achieve a fast forwarding ratio of $R^{FF}_{0.8} = 174$. 


\subsection{Randomized Training}\label{ap:LargeNeig}


While the fsVFF cost as stated in Eq.~\eqref{eq:fsVFFcost} has $n_{\rm eig}$ terms, this does not necessarily mean that the number of circuits required to evaluate it also scales with $n_{\rm eig}$. Analogous to mini-batch gradient descent methods popular for the training of classical neural networks, we can use only a small random selection of the total training dataset per gradient evaluation, yet over the whole optimization the total training set will be fully explored many times over. Therefore, instead of restricting ourselves to a discrete set of training states, which requires setting the size of the training set to be equal to or greater than $n_{eig}$, we can instead randomly select our training states from a continuum. This has the added advantage that it is then unnecessary to explicitly compute $n_{\rm eig}$.

This approach results in a modified cost function of the form 
\begin{equation}
    \widetilde{C}_{\rm fsVFF} := 1-\frac{1}{|R|} \sum\limits_{r \in R} |\bra{\psi_0} V(-r t_{\rm max}) U(rt_{\rm max}) \ket{\psi_0}|^2
\label{eq:randomised_cost}
\end{equation} where $V(t) = W D(t) W^\dagger$ is the fsVFF ansatz, $U(t)$ is a Trotter approximation for the short time unitary evolution, and the elements of the set $R$ are randomly generated numbers from the interval $[-1, 1]$. The gradients of the cost function are smaller when the unitary acts close to the identity operation, so to maintain stronger gradients it is advantageous for the elements of $R$ to be slightly biased towards the edges of the interval. Specifically, in our numerics to test this approach, the absolute magnitude of $r$ was raised to the power of 0.75. Although this approach does not require an {\it{a priori}} calculation of $n_{\rm eig}$, there is a caveat that $t_{max}$ needs to be large enough to get sufficient separation of the training states so they are not functionally identical. 
This alternative training setup is potentially a yet more NISQ friendly variant, as the unitary does not need to be decomposed into the form $U(\Delta t)^{n_{eig}}$, as required in Eq.~\ref{eq:fsVFFcost}, and therefore allows for shorter depth circuits. 

\begin{figure}[t!]
\centering
\includegraphics[width =\columnwidth]{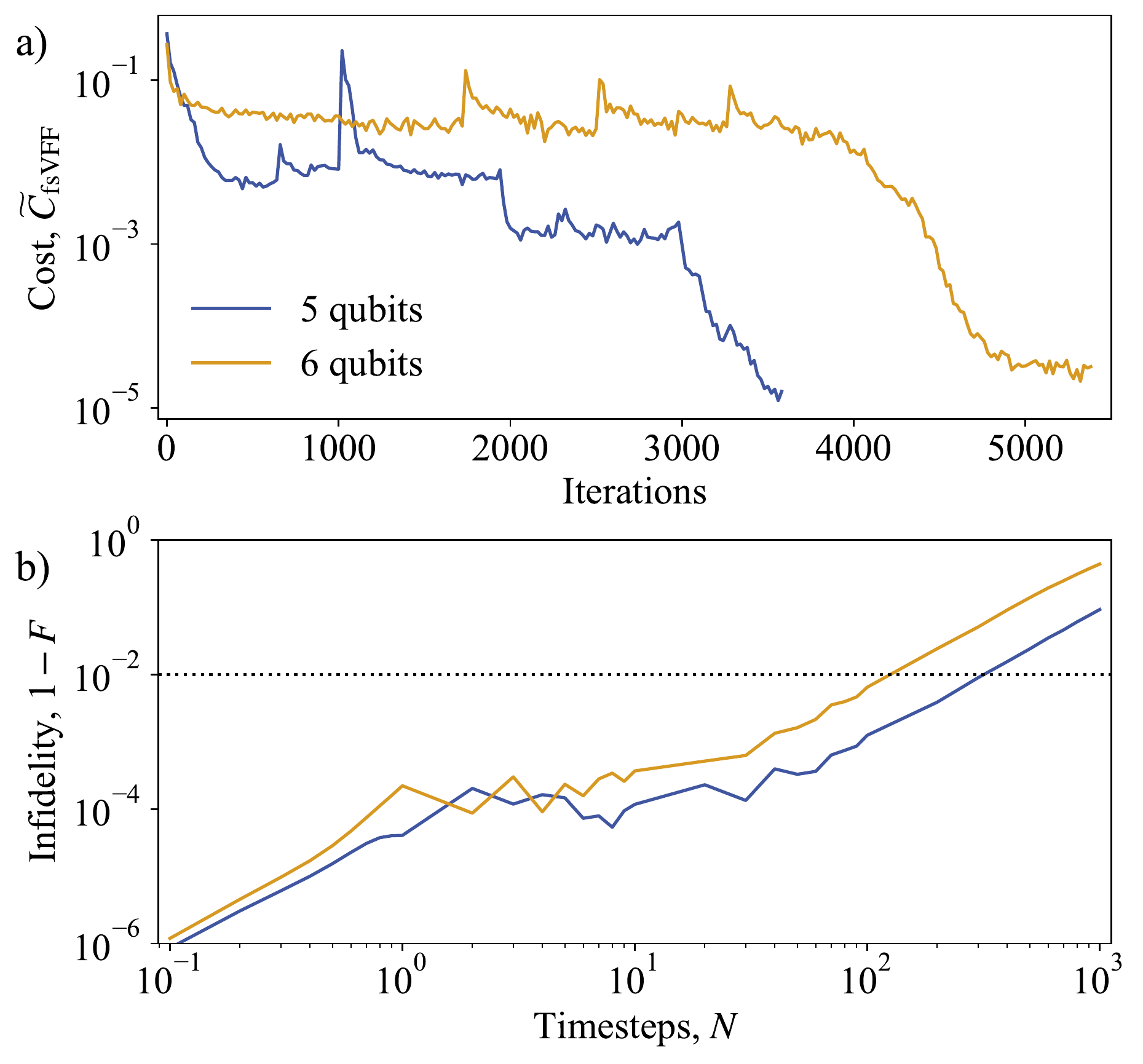}
\vspace{-2mm}
\caption{\textbf{Randomized Training.} a) The 5 (6) qubit XY Hamiltonian with initial state $\ket{11100} (\ket{111000})$ is diagonalized using the cost function Eq.~\eqref{eq:randomised_cost}, using only 2 training states per cost function evaluation to learn the evolution within the 9 (12) dimensional subspace. The final circuit found by the learning algorithm for the diagonalizing unitary, $W$, had 32 (134) CNOT gates.
b) After completion of the randomised training, the Hamiltonians were fast-forwarded, with the infidelity evaluated in comparison to the noiseless Trotter-iterated state $U(\Delta t)^N \ket{\psi_0}$. }
\label{fig:BatchedTraining}
\end{figure}
 
To demonstrate the viability of this batched training method we diagonalized the 5 (6) qubit XY Hamiltonian with initial state $\ket{11100} (\ket{111000})$, which has an $n_{eig} = 9 \ (12)$. For both training curves shown in Fig.~\ref{fig:BatchedTraining}, only 2 training states per cost function evaluation were used.
In both cases, we trained with the unitary $U(t/6)^6$ where $U$ was the second order Trotter-Suzuki operator, and $t_{max} = 1$.
The cost was successfully minimised to $10^{-5}$ in both cases, and a noiseless simulation error of less than $10^{-2}$ was maintained for over 100 time steps on fast forwarding.

\section{Energy estimation}

\begin{figure*}[t]
\includegraphics[width =2\columnwidth]{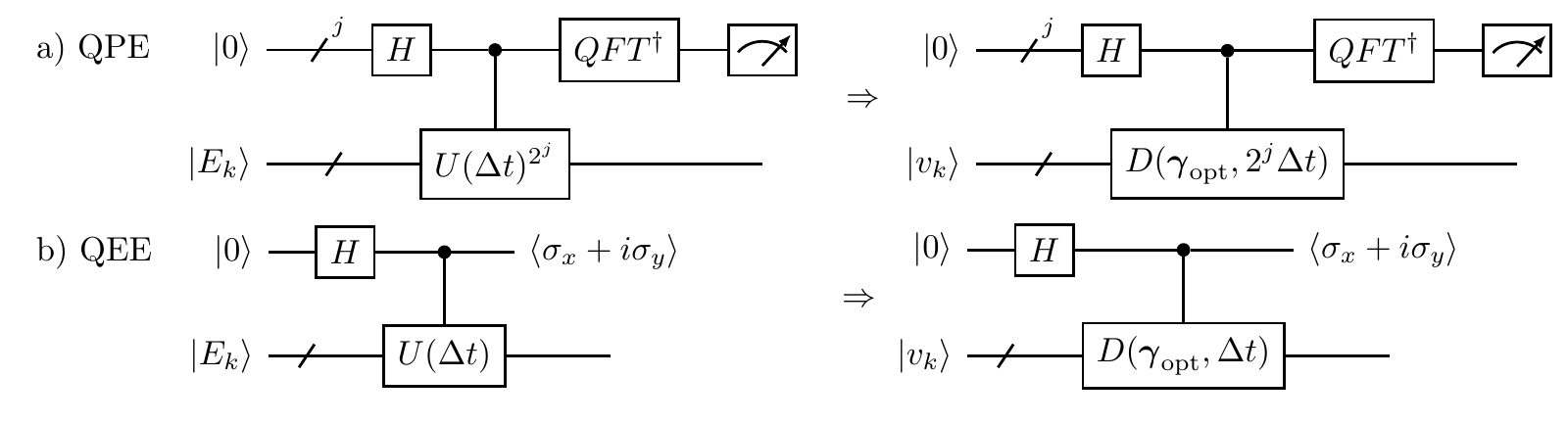}
\caption{\textbf{Energy estimation circuits} a)/b) show circuit diagrams depicting the enhancement of QPE/QEE using fsVFF. A circuit depth reduction is achieved through replacing $U(\Delta t)$ with $D(\gamma_{\rm opt}, \Delta t)$, and removing the need to prepare an eigenstate in favour of a computational basis state, $\ket{v_k} = W^{\dagger}\ket{E_k}$. QPE relies on implementing controlled unitaries of the form $U(\Delta t)^{2^j}$ and therefore replacing these with $D(\gamma_{\rm opt}, 2^j \Delta t)$ results in an exponential reduction in circuit depth.}
\label{fig:QPE_circuits}
\end{figure*}

The diagonalization obtained from the optimization stage of fsVFF, $ W(\thv_{\rm opt}) D(\gav_{\rm opt}, \Delta t) W(\thv_{\rm opt})^\dagger$, implicitly contains approximations of the eigenvalues and eigenvectors of the Hamiltonian of the system of interest. In this section we discuss methods for extracting the energy eigenstates and eigenvalues from a successful diagonalization and implement them on quantum hardware. 

The energy eigenvectors spanned by the initial state $\ket{\psi_0}$ can be determined by the following simple sampling method. The first step is to apply $W^\dagger$ to the initial state $\ket{\psi_0}$. In the limit of perfect learning and vanishing Trotter error, this gives
\begin{equation}
    W(\thv_{\rm opt})^{\dagger}\ket{\psi_0} = \sum_{k=1}^{n_{eig}} a_k \ket{v_k} 
\end{equation}
where $a_k = \langle E_k| \psi_0 \rangle$ and $\{\ket{v_k}\}_{k=1}^{n_{eig}}$ is a set of computational basis states.
The energy eigenstates spanned by $\ket{\psi_0}$ are then found by applying $W(\thv_{\rm opt})$ to any of the states obtained from measuring $W(\thv_{\rm opt})^{\dagger}\ket{\psi_0}$ in the computational basis, that is $\{ \ket{E_k} \}_{k = 1}^{n_{\rm eig}} = \{ W(\thv_{\rm opt}) \ket{v_k} \}_{k = 1}^{n_{\rm eig}}$. 

Extracting the energy eigenvalues from $D$ is more subtle. Firstly, as $W D W^\dagger$ and $U$, even in the limit of perfectly minimizing the cost $C_{\rm fsVFF}$, may disagree by a global phase $\phi$, at best we can hope to learn the difference between, rather than absolute values, of the energy eigenvalues of $H$. For simple cases, where the diagonal ansatz $D$ is composed of a polynomial number of terms, these energy value differences may be extracted directly by rewriting $D$ in the computational basis. For example, in our hardware implementation $D(\gamma) = \exp\left(- i \frac{\gamma \Delta t Z_1}{2} \right) \otimes \id$ and therefore the difference in energy between the two eigenvalues spanned by $\ket{\psi_0}$ is given by $\gamma_{\rm opt} + \frac{k \pi}{\Delta t}$. Here $k$ is an integer correcting for the arbitrary phase arising from taking the $\log$ of $D$ that can be determined using the method described in Ref.~\cite{commeau2020variational}. Using this approach, we obtain 1.9995 and 2.0019 from the training on IBM and Rigetti respectively, in good agreement with the theoretically expected value of~2. For more complex cases, this simple post-processing method will be become intractable and an algorithmic approach will be necessary.

\begin{figure}[t]
\centering
\includegraphics[width =\columnwidth]{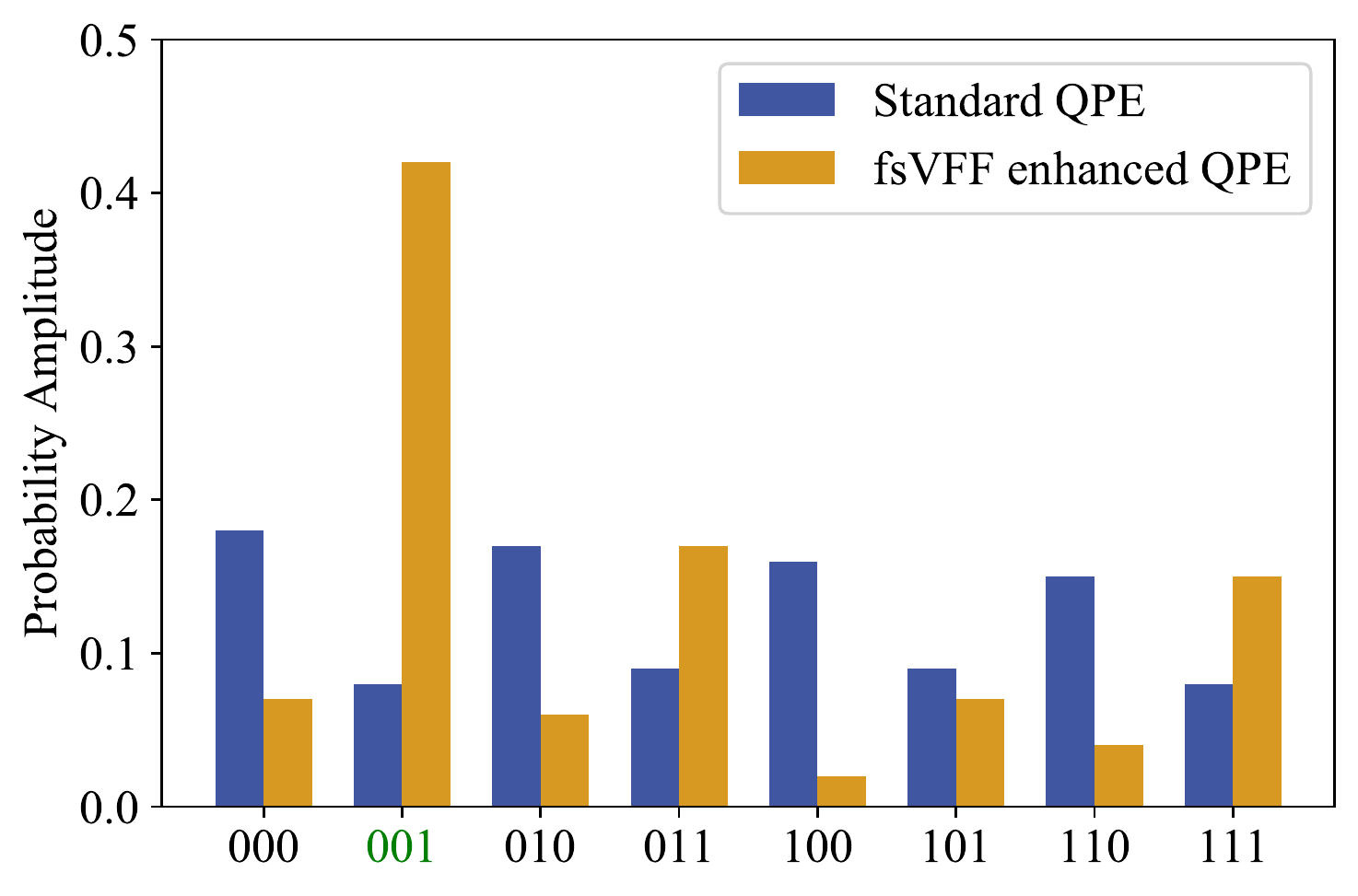}
\vspace{-2mm}
\caption{\small \textbf{Quantum Phase Estimation.} Using the 2-qubit diagonalization found from training on ibmq\_toronto, QPE was performed on ibmq\_boeblingen on the eigenvector $\ket{E_1} := \frac{1}{\sqrt{2}}(\ket{01}+\ket{10})$. A phase of $e^{\frac{2\pi i}{8}}$ is applied, so the measured output should be 001 with probability 1. The variation distance from the target probability distribution when using QPE with fsVFF was 0.578, compared to 0.917 using standard QPE.}
\label{fig:QPEimplementation}
\end{figure}

Quantum Phase Estimation (QPE)~\cite{nielsen2000quantum} and Quantum Eigenvalue Estimation (QEE)~\cite{somma2020quantum} are fault tolerant quantum algorithms for estimating the eigenvalues of a unitary operation. However, their implementation on current quantum devices is limited by the reliance on the execution of controlled unitaries from ancillary qubits. These controlled unitaries require many entangling gates, and introduce too much noise to be realized for large scale systems on current hardware. Once an evolution operator has been diagonalized in the subspace of an initial state, fsVFF can be used to significantly reduce the circuit depth of QPE and QEE, as shown in Figure~\ref{fig:QPE_circuits}. In this manner, fsVFF provides a NISQ friendly means of estimating the eigenvalues within a subspace of a Hamiltonian.

To demonstrate the power of fsVFF to reduce the depth of QPE, we perform QPE using the diagonalization obtained from training on IBM's quantum computer. Specifically, we consider the input eigenvector $\ket{E_1} := \frac{1}{\sqrt{2}}(\ket{01}+\ket{10})$. This is one of the eigenvectors spanned by the input state of our earlier hardware implementation, $\ket{\psi_0} = \ket{10}$. We then consider evolving $\ket{E_1}$ under $H_{\rm XY}$ for a time step of $\Delta t = 1/8$. Since the energy of the state $\ket{E_1}$ equals 1, we expect this to result in a phase shift of $e^{2\pi i/8}$ being applied to $\ket{E_1}$.
We implemented QPE and fsVFF enhanced QPE to measure this phase using the circuits shown in Fig~\ref{fig:QPE_circuits}. We chose to measure to 3 bits of precision and therefore the output should be the measurement 001 with probability one. As Figure~\ref{fig:QPEimplementation} shows, it appears that the standard QPE implementation was unable to discern this phase. In contrast, when fsVFF was used to reduce the circuit depth, the output distribution was strongly peaked at the correct state. 

\begin{figure}[t]
\includegraphics[width =\columnwidth]{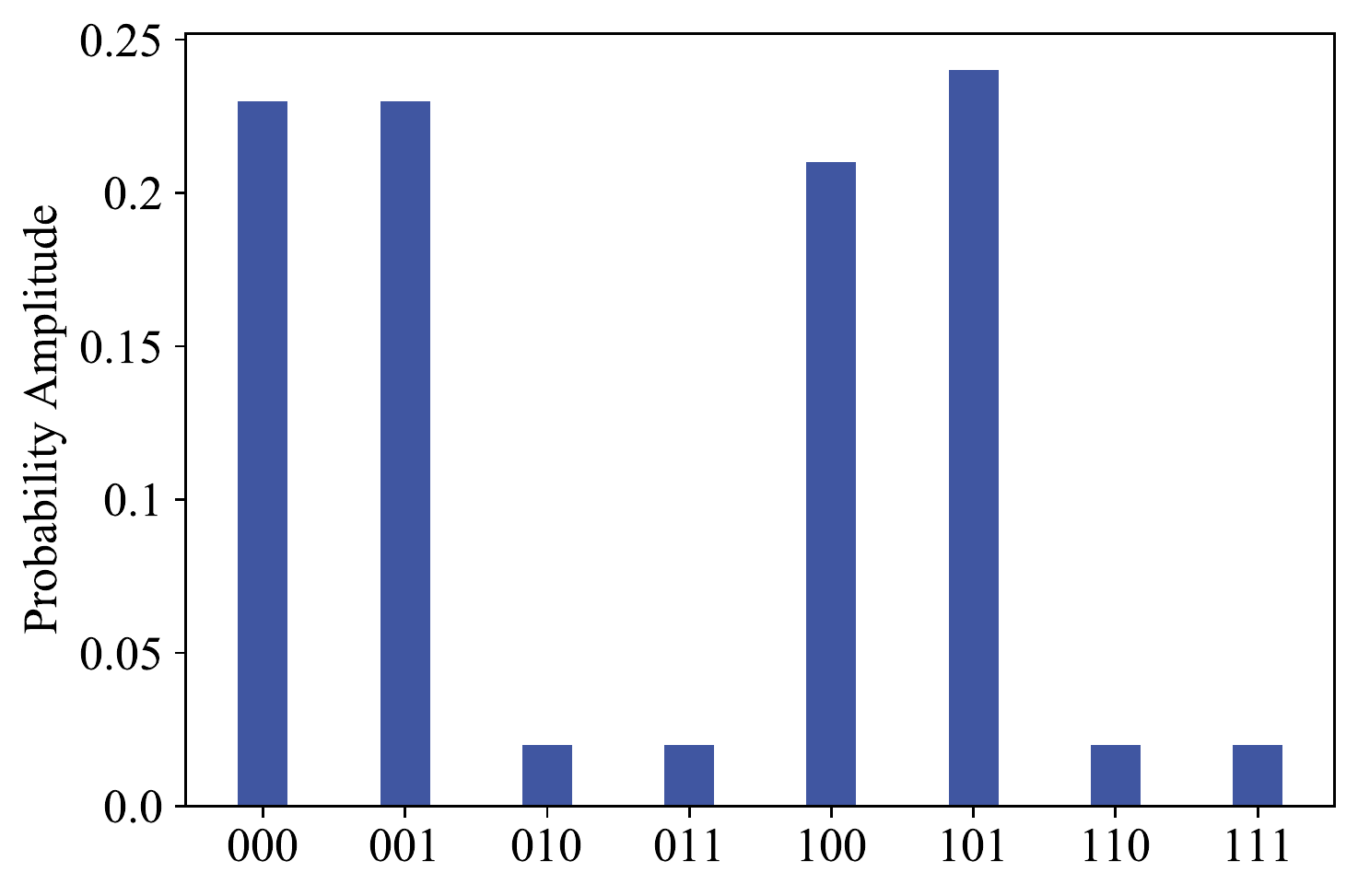}
\vspace{-2mm}
\caption{\textbf{Determining the eigenstates spanned by the initial state}: The 3-qubit XY Hamiltonian was diagonalized on a quantum simulator in the subspace of initial state $\ket{\psi_0} = \ket{110}$ to obtain $\thv_{
\rm opt}$ and $\gav_{\rm opt}$. Here we show the output of measuring $W(\thv_{\rm opt})^{\dagger} \ket{\psi_0}$ in the computational basis on ibmq\_boeblingen. The 4 non-zero states correspond to the 4 eigenvectors spanned by $|\psi_0\rangle$.}
\label{fig:QEE_comp_basis}
\end{figure}

\begin{figure}[t]
\includegraphics[width=58mm]{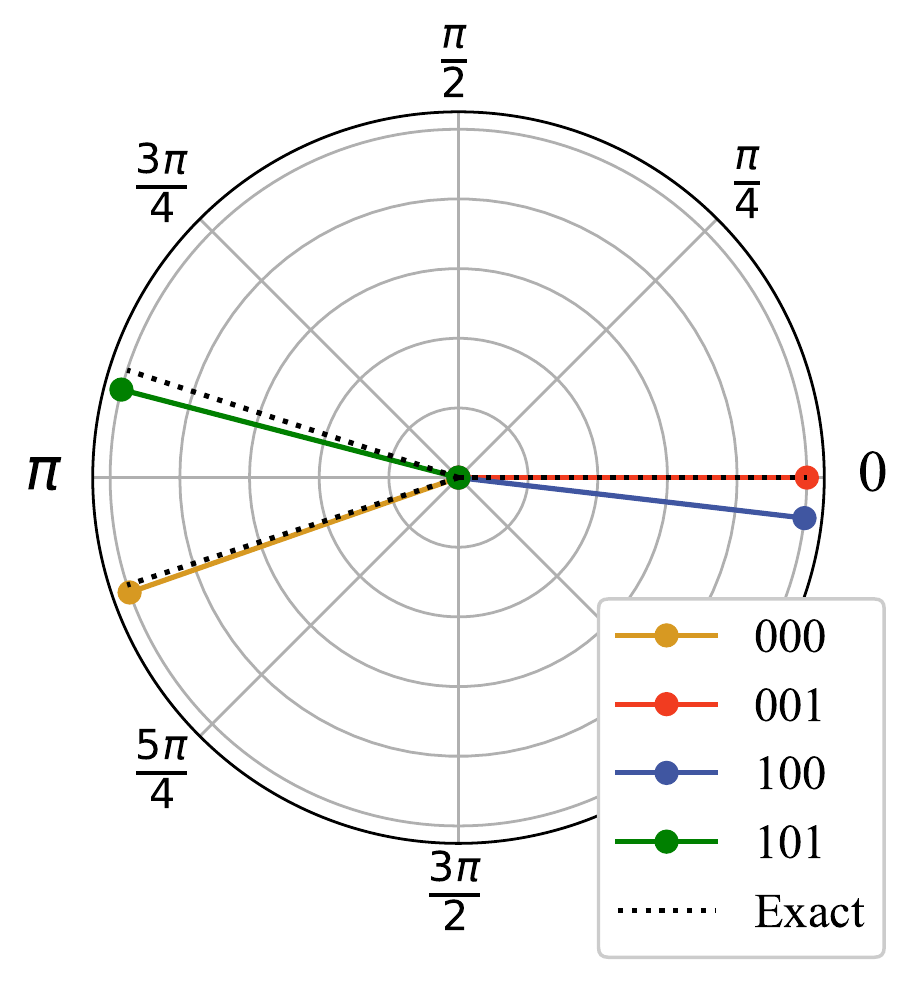}
\vspace{-2mm}
\caption{\textbf{Eigenvalue Estimation using QEE.} Here we show the result of implementing QEE (using fsVFF as a pre-processing step) on ibmq\_santiago to calculate the eigenvalues of the eigenvectors in the subspace spanned by $\ket{011}$. The solid yellow, red, blue and green lines represent the eigenvalues obtained for the $\ket{000}$, $\ket{001}$, $\ket{100}$ and $\ket{101}$ states, with exact corresponding energies of \{-2.828, 0, 0, 2.828\}, indicated by the dotted lines. The eigenvalues are plotted as phases since for $\Delta t = 1$ there is a one to one correspondence.}
\label{fig:QEE_eigenvalues}
\end{figure}


QEE requires only one ancillary qubit, a single implementation of $e^{-iHt}$, and no Quantum Fourier Transform and therefore is less resource intensive than QPE. Nonetheless, we can again, as shown in Fig.~\ref{fig:QPE_circuits}, use fsVFF as a pre-processing step to reduce the circuit depth. 

We tested this on the 3-qubit XY Hamiltonian by first performing fsVFF on a quantum simulator with the initial state $\ket{\psi_0} = \ket{110}$. Having obtained an approximate diagonalization, we determined the eigenstates using the sampling method described earlier. Figure~\ref{fig:QEE_comp_basis} shows the results of the measurement of $W(\thv_{\rm opt})^\dagger \ket{\psi_0}$, with four strong peaks corresponding to the four eigenvectors in this subspace. 

Figure~\ref{fig:QEE_eigenvalues} shows the results of QEE implemented on ibmq\_boeblingen. We use the basis states found from the sampling method as our inputs to reduce the depth of the circuit, and remove the need to use the time-series method originally proposed for extracting the eigenvalues, as we could calculate the eigenvalues individually by inputting their corresponding eigenvectors. A value of $\Delta t = 1$ was used so the phase calculated directly matched the eigenvalue. After removing a global phase, QEE had accurately found the eigenvalues of the four eigenvectors, with a mean-squared error from the true values of $4.37 \times 10^{-3}$. 





\section{Discussion}

In this work, we demonstrated that despite the modest size and noise levels of the quantum hardware that is currently available, it is possible to perform long time dynamical simulations with a high fidelity. Specifically, we have introduced fsVFF, a new algorithm for NISQ simulations, which we used to simulate a 2-qubit XY-model spin chain on the Rigetti and IBM quantum computers. We achieved a fidelity of at least 0.9 for over 600 time steps. This is a 150-fold improvement on the standard iterated Trotter approach, which had a fidelity of less than 0.9 after only 4 time steps. Moreover, our numerical simulations of the 4 qubit XY model and 8 qubit Fermi-Hubbard model achieved fast-forwarding ratios of 87.5 and 174 respectively, indicating the viability of larger implementations in the near future as hardware improves.

Central to the success of the fsVFF algorithm is the fact that it is tailored to simulating a particular fixed initial state rather than an arbitrary initial state. By focusing on this less demanding task, we showed that it is possible to substantially reduce the width and depth of the previously proposed VFF algorithm. In particular, since fsVFF only requires finding a diagonalization of a short-time evolution unitary on the subspace spanned by the initial state (compared to the entire Hilbert space in the case of VFF), fsVFF can utilize much simpler ans\"{a}tze. This is demonstrated in our hardware implementation, where one CNOT and two parameterized single qubit rotations proved sufficient for an effective ansatz for $W$ and one single qubit rotation was sufficient for $D$. 


The fsVFF algorithm, similarly to VFF, is fundamentally limited by the initial Trotter error approximating the short time evolution of the system. The Variational Diagonalization Hamiltonian (VHD) algorithm~\cite{commeau2020variational} may be used to remove this error. However, like VFF, VHD is designed to simulate any possible initial state. There are a number of different approaches inspired by fsVFF that could be explored for reducing the resource requirements of the VHD algorithm by focusing on simulating a particular initial state. Such a ``fixed state VHD'' algorithm would allow for more accurate long time simulations on NISQ hardware. 

More generally, our work highlights the trade off between the universality of an algorithm and the resources required to implement it. One can imagine a number of alternative ways in which the universality of an algorithm can be sacrificed, without significantly reducing its utility, in order to make it more NISQ friendly. For example, one is often interested in studying the evolution of a particular observable of interest, rather than all possible observables. It would be interesting to investigate whether a fixed-observable fsVFF could further reduce the resources required to implement long time high fidelity simulations. More broadly, an awareness of this trade off may prove useful beyond dynamical simulation for the ongoing challenge of adapting quantum algorithms to the constraints of NISQ hardware.

\section{Acknowledgements}

  JG and KG acknowledge support from the U.S. Department of Energy (DOE) through a quantum computing program sponsored by the Los Alamos National Laboratory (LANL) Information Science \& Technology Institute. ZH, BC and PJC acknowledge support from the LANL ASC Beyond Moore's Law project. ZH acknowledges subsequent support from the Mark Kac Fellowship. We acknowledge the LANL Laboratory Directed Research and Development (LDRD) program for support of AS and initial support of BC under project number 20190065DR as well as LC under project number 20200022DR. AA was supported by the U.S. Department of Energy (DOE), Office of Science, Office of High Energy Physics QuantISED program under Contract No.~DE-AC52-06NA25396. LC and PJC were also supported by the U.S. DOE, Office of Science, Basic Energy Sciences, Materials Sciences and Engineering Division, Condensed Matter Theory Program. This research used quantum computing resources provided by the LANL Institutional Computing Program, which is supported by the U.S. Department of Energy National Nuclear Security Administration under Contract No. 89233218CNA000001. This research used additional quantum  computational resources supported by the LANL ASC Beyond Moore's Law program and by the Oak Ridge Leadership Computing Facility, which is a DOE Office of Science User Facility supported under Contract DE-AC05-00OR22725.




\vfill
\bibliography{quantum.bib}
\onecolumngrid

\appendix

\clearpage

\section{Faithfulness of cost function}\label{ap:Faithful}


Here we demonstrate that the cost function is faithful in the limit that leakage from the subspace spanned by the initial state can be disregarded. That is, suppose we could compile the exact evolution of the system $U = \exp(- i H \Delta t)$ for a short timestep $\Delta t$. Then the cost function vanishes,
\begin{equation}\label{eq:CostAP}
    C_{\rm fsVFF}(U,V, \psi_0) = 1 - \frac{1}{n_{\rm eig}} \sum_{k=1}^{n_{\rm eig}} | \bra{\psi_0} {V^\dagger}^k U^k \ket{\psi_0} |^2 = 0 \, ,
\end{equation}
if and only if the fidelity of the fast-forwarded simulation is perfect,
\begin{equation}
  F_\tau =  | \bra{\psi_0} {V^\dagger}^\tau U^\tau \ket{\psi_0} |^2 = 1 \, ,
\end{equation}
for all times $\tau$. Note, the reverse direction is trivial. If $F_\tau =1$ for all $\tau$ then $C_{\rm fsVFF}(U,V, \psi_0) = 0$. 

\medskip

Before embarking on the core of the proof let us first emphasize that in the definition of the cost \eqref{eq:CostAP} we average over $n_{\rm eig}$ terms, where $n_{\rm eig}$ is the number of eigenstates spanned by the initial state $\ket{\psi_0}$ corresponding to unique eigenvalues of the Hamiltonian $H$. The restriction to eigenstates with \textit{unique} eigenvalues is important since it is this which determines the dimension of the subspace spanned by the future evolution of $\ket{\psi_0}$ (which in turn determines the number of training states required to learn $U = \exp(- i H \Delta t)$). 

To see this consider a Hamiltonian $H$ with a spectrum $\{E_k\}_{k=1}^{2^n}$ and corresponding eigenstates $\{ \ket{E_k} \}_{k=1}^{2^n}$. The initial state $\ket{\psi_0}$ can be expanded in the energy eigenbasis as 
\begin{equation}\label{eq:StateExpan}
    \ket{\psi_0} = \sum_{k=1}^{m} a_k \ket{E_k} \, .
\end{equation}
The future evolution of such a state, i.e. the set of states
\begin{equation}
    \mathcal{S}_k (\psi_0 , H) := \{ e^{- i H j \Delta t} \ket{\psi_0} \}_{j = 0}^{k} \, 
\end{equation}
with $k = \infty$,
is solely contained within the subspace spanned by $\{ \ket{E_k} \}_{k=1}^{m}$ since 
\begin{equation}
    e^{- i H j \Delta t} \ket{\psi_0} = \sum_{k=1}^{m} a_k e^{- i E_k j \Delta t}  \ket{E_k} \, .
\end{equation}
If the spectrum $\{E_k\}_{k=1}^{2^n}$ is non-degenerate then the evolution generates relative phases between all the $m = n_{\rm eig}$ eigenstates spanned by $\ket{\psi_0}$. In that case, $\mathcal{S}_\infty$ will span the entirety of the $m$ dimensional subspace spanned by $\{ \ket{E_k} \}_{k=1}^{m}$, i.e., an $m = n_{\rm eig}$ dimensional space.
However, suppose the eigenstate expansion of $\ket{\psi_0}$ includes two degenerate eigenstates, i.e. two eigenstates that share the same eigenvalue. In that case the evolution generates relative phases between only $m - 1$ of the eigenstates $\{ \ket{E_k} \}_{k=1}^{m}$ and therefore $\mathcal{S}_\infty$ is confined to an $m - 1$ dimensional subspace. More generally, if the eigenstate expansion of $\ket{\psi_0}$ includes $n_{\rm eig}$ states with unique eigenvalues, then evolution under $U = \exp(- i H \Delta t)$ generates relative phases between $n_{\rm eig}$ states and so $\mathcal{S}_\infty$ will span an $n_{\rm eig}$ dimensional subspace of the space. In this manner, it is the number of eigenstates spanned by the initial state $\ket{\psi_0}$ corresponding to \textit{unique} eigenvalues, $n_{\rm eig}$, that determines the subspace spanned by $\mathcal{S}_\infty$. 

We note that any initial state $\ket{\psi_0}$ can be written in the form Eq.~\eqref{eq:StateExpan} where, crucially, the sum is over only eigenstates corresponding to unique energies, i.e. $m = n_{\rm eig}$.
For a Hamiltonian with a non-degenerate spectrum this expansion is trivial. For a Hamiltonian with a degenerate spectrum there is some freedom in how the eigenstates corresponding to degenerate eigenvalues are defined, since the superposition of degenerate eigenstates is also an eigenstate corresponding to the same energy. Therefore, henceforth, for degenerate Hamiltonians, we suppose that the energy eigenbasis $\{ \ket{E_k} \}_{k=1}^{2^n}$ is defined such that the eigenstate expansion of $\ket{\psi_0}$ only contains eigenstates with unique energies, that is $m = n_{\rm eig}$ terms. 

We remark that our approach here may also be framed in the language of Krylov spaces. The Krylov subspace~\cite{krylov1931numerical} associated with the operator $U$ and vector $\ket{\psi_0}$ is the linear subspace spanned by the vectors generated by evolving $\ket{\psi_0}$ under $U$ up to $k$ times. That is
\begin{equation}
    \mathcal{K}_k(U, \psi_0) = \text{span} \{ \mathcal{V}_k  \} \ \ \ \ \text{where} \ \ \ \ \mathcal{V}_k := \{ \ket{\psi_l} \}_{l=0}^{l=k} \, ,
\end{equation}
with $ \ket{\psi_l} := U^l \ket{\psi_0}$. Now supposing $U = e^{-i H \Delta t}$ and   $\ket{\psi_0} = \sum_{k=1}^{n_{\rm eig}} a_k \ket{E_k}$, we have that $\mathcal{V}_k = \mathcal{S}_k$. Thus the future evolution of $\ket{\psi_0}$ is confined to the Krylov space $\mathcal{K}_\infty(U, \psi_0)$. This is an $n_{\rm eig}$ dimensional subspace, and therefore, as will become clear, we require $n_{\rm eig}$ training states in order to learn $U$ on this subspace.

\medskip

To prove the forward direction, we first note that if $C_{\rm fsVFF} = 0$ then as $ 0 \leq | \bra{\psi_0} {V^\dagger}^k U^k \ket{\psi_0} |^2 \leq 1$, we have that $| \bra{\psi_0} {V^\dagger}^k U^k \ket{\psi_0} |^2 = 1$ for all $k$. It thus follows that the action of $V^k$ on $\ket{\psi_0}$ agrees with the action of $U^k$ on $\ket{\psi_0}$ up to an unknown phase $e^{i\phi_k}$, i.e. for $ 1 \leq k \leq n_{\rm eig} $ we have that
\begin{align}\label{eq:V-on-traindata}
V^k \ket{\psi_0} =e^{i\phi_k} U^k \ket{\psi_0}   \, .
\end{align}
Or equivalently, the action of $V$ and $U$ agree on the training states $\SC_{\rm train} := \{ \ket{\psi_k} \}_{k=0}^{n_{\rm eig} -1 }$ up to an unknown phase $e^{i\tilde{\phi}_k}$, that is
\begin{align}\label{eq:PsiKAgree}
V \ket{\psi_k} =e^{i\tilde{\phi}_k} U \ket{\psi_k} \, \forall \, \ket{\psi_k} \, \in \, \SC_{\rm train},
\end{align}
where $\tilde{\phi_k} = \phi_{k+1} - \phi_k$.

Now by construction (see Section~\ref{sec:Neig}) the $n_{\rm eig}$ training states $\SC_{\rm train}$ are linearly independent. Furthermore, since the initial simulation time $\Delta t$ may be chosen freely, the unitary $U =  \exp(- i H \Delta t)$ can be chosen such that none of the states in $\SC_{\rm train}$ are orthogonal. In this case, the unknown phases all agree and we have that $\tilde{\phi_k} = \phi$ for all $k$. To see this note that given $\ket{\psi_1}$ and $\ket{\psi_0}$ are linearly independent but non-orthogonal, the state $\ket{\psi_1}$ can be represented as follows:
\begin{align}
    \ket{\psi_1} = c \ket{\psi_0} + c_\perp \ket{\psi_0^{\perp}},
\end{align}
where $|c|^2  + |c_\perp|^2  = 1$ and $|c|^2 \neq 0$. Then from \eqref{eq:PsiKAgree}, we find that 
\begin{align}
    e^{-i \tilde{\phi}_1} &= \langle \psi_1 \vert W \vert \psi_1 \rangle\\
    & = \vert c \vert^2 e^{i \tilde{\phi_0}} + \vert c_\perp \vert^2 \langle \psi_0^{\perp} \vert W \vert \psi_0^{\perp}\rangle \\
    &= \vert c \vert^2 e^{i \tilde{\phi_0}} + (1- \vert c \vert^2 )\langle \psi_0^{\perp} \vert W \vert \psi_0^{\perp}\rangle  \, ,
\end{align}
where $W := V^\dagger U$ (note that this use of $W$ is distinct from $W(\thv)$ used in the main text for a parameterized eigenvector unitary). The above expression can be rearranged as 
\begin{equation}
    e^{-i \tilde{\phi}_1} - \langle \psi_0^{\perp} \vert W \vert \psi_0^{\perp}\rangle = \vert c \vert^2 ( e^{-i \tilde{\phi}_0} -  \langle \psi_0^{\perp} \vert W \vert \psi_0^{\perp}\rangle ) \, .
\end{equation}
Since $|e^{i \theta_1}|=1$ and $|c|^2 \neq 0$ the aforementioned equation is satisfied if and only if 
\begin{align}
\langle \psi_0^{\perp} \vert W \vert \psi_0^{\perp}\rangle=  e^{-i \tilde{\phi}_1} =  e^{-i \tilde{\phi}_0} \, ,
\end{align}
which implies that $\tilde{\phi}_0 = \tilde{\phi}_1$. Then by recursively applying this procedure to the rest of the states in the training set, we find that $\tilde{\phi}_k = \tilde{\phi}_j:= \phi$ for $ 0 \leq k \leq n_{\rm eig} -1  $ and  $ 0 \leq j \leq n_{\rm eig} -1  $ .

To understand the constraints from the minimization of the cost function, it is helpful to consider the form of the unitary matrix $W$. It follows from Eq.~\eqref{eq:PsiKAgree}, and the fact that since $W$ is unitary ($\sum_j \vert W_{ij}\vert^2 = \sum_i \vert W_{ij}\vert^2 =1 $), that $W$ can be represented as~\cite{poland2020no}:
\[
W =\left(\begin{array}{@{}c|c@{}}
   \begin{matrix}
  e^{i\phi} & \dots & 0 \\
  \vdots & \ddots & \\
  0 &   & e^{i \phi}
  \end{matrix}
  & \bigzero \\
\hline
  \bigzero &
  W_{\perp} \\
\end{array}\right) .
\] 
Here the upper left hand block spans the $n_{\rm eig}$ dimensional subspace spanned by the input training states $\SC_{\rm train}$ and $W_{\perp}$ is an unknown unitary matrix acting on the $(d- n_{\rm eig})$ dimensional space orthogonal to $\SC_{\rm train}$. For later convenience let us also note that the matrix $\tilde{W}:=  U V^\dagger$ is also a unitary matrix of the form 
\[
\tilde{W} =\left(\begin{array}{@{}c|c@{}}
   \begin{matrix}
  e^{i\phi} & \dots & 0 \\
  \vdots & \ddots & \\
  0 &   & e^{i \phi}
  \end{matrix}
  & \bigzero \\
\hline
  \bigzero &
  \tilde{W}_{\perp} \\
\end{array}\right),
\] 
where $\tilde{W}_{\perp}$ is again a $(d- n_{\rm eig})$ dimensional unitary matrix.

We are now in a position to show that if $C_{\rm fsVFF}(U,V, \psi_0) = 0$, and by construction the states in $\SC_{\rm train}$ are linearly independent and non-orthogonal, then 
\begin{equation}
  F_\tau =  | \bra{\psi_0} {V^\dagger}^\tau U^\tau \ket{\psi_0} |^2 = 1 \, ,
\end{equation}
for all times $\tau$. To see this first note that for all times $\tau \leq n_{\rm eig}$ that $F_\tau =1$ follows directly from Eq.~\eqref{eq:V-on-traindata}. Now by construction, any state $\ket{\psi_\tau}$ for $\tau > n_{\rm eig}$ linearly depends on the states in $\SC_{\rm train}$ and so it can be written as
\begin{equation}
    \ket{\psi_\tau} = \sum_{j=0}^{n_{\rm eig}}  b^{(\tau)}_j \ket{\psi_j} \, ,
\end{equation}
where $b^{(\tau)}_j = \braket{\psi_j|\psi_\tau}$. Thus we have that 
\begin{equation}\label{eq:W-on-psi}
    W \ket{\psi_\tau} =  \tilde{W} \ket{\psi_\tau} = e^{i \phi}  \ket{\psi_\tau}
\end{equation}
for any time $\tau$. It straightforwardly follows from  Eq.~\eqref{eq:V-on-traindata} and Eq.~\eqref{eq:W-on-psi} that the simulation fidelity at time $\tau = n_{\rm eig}+1$ equals 1,
\begin{align}
    F_{n_{\rm eig}+1}  &=  | \bra{\psi_0} {V^\dagger}^{n_{\rm eig}+1} U^{n_{\rm eig}+1} \ket{\psi_0} |^2 \\ &= | \bra{\psi_{n_{\rm eig}}} W \ket{\psi_{n_{\rm eig}}} |^2  = 1 \, .
\end{align}
Now, let us consider the simulation fidelity at time $\tau = n_{\rm eig}+2$, \begin{align}
    F_{n_{\rm eig}+2}  &=  | \bra{\psi_0} {V^\dagger}^{n_{\rm eig}+2} U^{n_{\rm eig}+2} \ket{\psi_0} |^2 \\
    &=  | \bra{\psi_{n_{\rm eig}}} V^\dagger W U \ket{\psi_{n_{\rm eig}}} |^2 \\
    &= | \bra{\psi_{n_{\rm eig}}} U^\dagger \tilde{W} W U \ket{\psi_{n_{\rm eig}}} |^2    \\ 
    &= | \bra{\psi_{n_{\rm eig}+1}}  \tilde{W} W \ket{\psi_{n_{\rm eig}+1}} |^2  = 1 \, , 
\end{align}
where we have again used Eq.~\eqref{eq:V-on-traindata} and Eq.~\eqref{eq:W-on-psi}.
Finally, the simulation fidelity at an arbitrary time $\tau > n_{\rm eig}$ can be evaluated as follows,
\begin{equation}
     F_\tau =  | \bra{\psi_0} {V^\dagger}^\tau U^\tau \ket{\psi_0} |^2 
     = | \bra{\psi_0} (U^\dagger \tilde{W})^\tau U^\tau \ket{\psi_0} |^2 = 1 \, .
\end{equation}
Thus, as claimed, if the cost function $C_{\rm fsVFF}$ vanishes the simulation fidelity is perfect for all times. 

\section{Noise Resilience of Cost}\label{ap:NoiseResilient}

To make a connection with prior results on noise resilience, we first note that $C_{\rm fsVFF}$, Eq.~\eqref{eq:fsVFFcost}, can be rewritten as 
\begin{equation}
      C_{\rm fsVFF}(U,V, \psi_0) =  \frac{1}{n_{\rm eig}} \sum_{k=1}^{n_{\rm eig}} C_{\rm fsVFF}^{(k)}(U,V, \psi_0) \, ,
\end{equation}
with 
\begin{equation}
    C_{\rm fsVFF}^{(k)}(U,V, \psi_0) =  1- \Tr\left[ |\mathbf{0} \rangle \langle \mathbf{0}| Y_k^{\tiny \psi_0}  | \mathbf{0} \rangle \langle \mathbf{0}| \left(Y_k^{\tiny \psi_0} \right)^\dagger  \right] \, .
\end{equation}
Here $Y^{\psi_0}_k :=  V_{\psi_0}^\dagger W D^k W^\dagger U^k V_{\psi_0}$ with $ V_{\psi_0}  \ket{\mathbf{0}} = \ket{\psi_0}$. Let $QC_k$ denote the circuit used to evaluate the cost term $C_{\rm fsVFF}^{(k)}$. Let $\tilde{C}_{\mbox {\tiny fsVFF}}^{(k)}$ denote the noisy version of the cost term $C_{\rm fsVFF}^{(k)}$, that is the cost evaluated when the circuit $QC_j$ is run in the presence of noise. Let $\mathbb{V}_k^{\rm opt}$ and $\mathbb{\tilde{V}}_k^{\rm opt}$ denote the sets of unitaries that optimize $C_{\rm fsVFF}^{(k)}$ and $\tilde{C}_{\mbox {\tiny fsVFF}}^{(k)}$ respectively. Now, it was shown in \cite{sharma2019noise} that the costs $C_{\rm fsVFF}^{(k)}$ are resilient to measurement noise, gate noise,
and Pauli channel noise in the sense that for circuits experiencing such noise we have $\mathbb{\tilde{V}}_k^{\rm opt} \subseteq \mathbb{V}_k^{\rm opt}$. This means that any set of parameters that minimize the noisy cost $\tilde{C}_{\mbox {\tiny fsVFF}}^{(k)}$ are guaranteed to also minimize the true exact cost $C_{\rm fsVFF}^{(k)}$. 

We will now argue that this implies that $ C_{\rm fsVFF}$ is also noise resilient. To do so, we first note that the costs $C_{\rm fsVFF}^{(k)}$ can be minimized simultaneously by any unitary $V^{\rm opt}$ that matches the target unitary $U$ up to a global phase $\phi$, i.e. such that $V^{\rm opt} = \exp(-i \phi) U$. Therefore, assuming that the ansatz for $V$ is sufficiently expressive that the costs $C_{\rm fsVFF}^{(k)}$ can be simultaneously minimized, the total cost $C_{\rm fsVFF}$ is minimized by the set of unitaries that minimize each of the $C_{\rm fsVFF}^{(k)}$ costs simultaneously, that is the set $ \cap_k \mathbb{V}_k^{\rm opt} := \mathbb{V}^{\rm opt}$. (Note, if the ansatz is not sufficiently expressive then it might not be possible to simultaneously minimize each of the terms and therefore the intersection might be empty). Now, given that $\mathbb{\tilde{V}}_k^{\rm opt} \subseteq \mathbb{V}_k^{\rm opt}$, it follows that $\mathbb{\tilde{V}}^{\rm opt} := \cap_k \mathbb{\tilde{V}}_k^{\rm opt}  \subseteq \mathbb{V}^{\rm opt}$. Thus, as claimed the total cost $C_{\rm fsVFF}$ is also noise resilient in the sense that any set of parameters that minimize the noisy cost $\tilde{C}_{\mbox {\tiny fsVFF}}$ also minimize the true exact cost $C_{\rm fsVFF}$.

\section{Local cost with trainability guarantee}\label{ap:LocalCost}

While $C_{\rm fsVFF}$ was motivated in Section~\ref{sec:Cost} as a natural choice of cost function to learn the evolution induced by a target unitary on a fixed initial state, it is expected to encounter what is known as a \textit{barren plateau} for large simulation sizes~\cite{mcclean2018barren,cerezo2020cost}. (See Refs.~\cite{cerezo2020impact,arrasmith2020effect,holmes2021connecting,volkoff2021large,sharma2020trainability,pesah2020absence,uvarov2020barren,marrero2020entanglement,patti2020entanglement, holmes2021barren}) for further details about the barren plateau phenomenon.) Namely, the gradient of the cost vanishes exponentially with $n$. As a result, for large systems the cost landscape is prohibitively flat and therefore an exponential precision is required to discern a minimization direction. This precludes successful training. 

In this appendix we introduce a local cost function to surmount this difficulty. To motivate our local cost, we first recall that $C_{\rm fsVFF}$, Eq.~\eqref{eq:fsVFFcost}, can be rewritten as 
\begin{equation}
   C_{\rm fsVFF}= 1 -  \frac{1}{n_{\rm eig}}\sum_{k=1}^{n_{\rm eig}} \Tr\left[ |\mathbf{0} \rangle \langle \mathbf{0}| Y_k^{\tiny \psi_0}  | \mathbf{0} \rangle \langle \mathbf{0}| \left(Y_k^{\tiny \psi_0} \right)^\dagger  \right] \, ,
\end{equation}
where $Y^{\psi_0}_k :=  V_{\psi_0}^\dagger W D^k W^\dagger U^k V_{\psi_0}$ with $ V_{\psi_0}  \ket{\mathbf{0}} = \ket{\psi_0}$. Analogously, we now define the local fixed state VFF cost as  
\begin{equation}
   C_{\rm fsVFF}^{\mbox {\tiny Local}}:= \frac{1}{n} \sum_{j=1}^{n}  C_{\rm fsVFF}^{\mbox {\tiny Local,j}}  \,  ,
\end{equation}
with
\begin{equation}
   C_{\mbox{\tiny fsVFF}}^{\mbox{\tiny Local,j}} = 1 -  \frac{1}{n_{\rm eig}} \sum_{k=1}^{n_{\rm eig}} \Tr\left[ \left( |0 \rangle \langle 0|_j \otimes \id_{\bar{j}} \right) Y_k^{\tiny \psi_0}  | \mathbf{0} \rangle \langle \mathbf{0}| \left(Y_k^{\tiny \psi_0} \right)^\dagger  \right] \, ,
\end{equation}
and where $\bar{j}$ denotes all qubits except the $j_{\rm th}$ qubit. Each of the $C_{\rm fsVFF}^{\mbox {\tiny Local, j}}$ terms can be measured using the same Loschmidt echo circuit as $C_{\rm fsVFF}$ but the final measurement is performed on just the $j_{\rm th}$ qubit (rather than all qubits).

Following the proof techniques of \cite{khatri2019quantum,sharma2019noise}, it is possible to show that 
\begin{equation}
    C_{\rm fsVFF}^{\mbox {\tiny Local}} \leq C_{\rm fsVFF} \leq n C_{\rm fsVFF}^{\mbox {\tiny Local}} \, .
\end{equation}
It therefore follows that $C_{\rm fsVFF}^{\mbox {\tiny Local}} = 0$ if and only if $C_{\rm fsVFF} = 0$. Thus, given that $C_{\rm fsVFF}$ is faithful (as shown in Appendix~\ref{ap:Faithful}), its local variant $C_{\rm fsVFF}^{\mbox {\tiny Local}}$ is also faithful.

Crucially, as $C_{\rm fsVFF}^{\mbox {\tiny Local}}$ is local, i.e. requires only local rather than global measurements, as long as the ansatz is not too deep, the cost landscape will be sufficiently featured for effective training. We therefore advocate using $C_{\rm fsVFF}^{\mbox {\tiny Local}}$ for simulations of larger systems.


\section{Cost function gradient derivation}\label{ap:Gradients}

Here we derive the analytic expressions for the gradient of the cost function $C_{\rm fsVFF}(U,V, \psi_0)$ for gradient descent optimisation. To emphasize the independence of each of the terms in the cost function and its dependence on $\thv$ and $\vec{\gamma}$ we write
\begin{equation}
      C_{\rm fsVFF}(U,V, \psi_0) =  \frac{1}{n_{\rm eig}} \sum_{k=1}^{n_{\rm eig}} C_{\rm fsVFF}^{(k)}(\vec{\theta}, \vec{\gamma}) \, ,
\end{equation}
where we have 
\begin{align}
 C_{\rm fsVFF}^{(k)}(\vec{\theta}, \vec{\gamma}) &:=  1 - | \bra{\psi_0} W(\vec{\theta}) D(k\gav) W(\vec{\theta})^\dagger U^k \ket{\psi_0} |^2  \\
 &=  1 - \Tr[X W(\vec{\theta}) D(k\gav) W(\vec{\theta})^\dagger Y_k W(\vec{\theta}) D(k\gav)^\dagger W(\vec{\theta})^\dagger]  \, ,
\end{align}
with $X =\ket{\psi_0} \bra{\psi_0} $ and
$Y_k = U^k \ket{\psi_0} \bra{\psi_0} (U^\dagger)^k $.

\paragraph*{Expressions.}

With the ansatz in \eqref{eqn:Vansatz}, the partial derivative of $C_{\rm fsVFF}(U,V, \psi_0)$ with respect to $\theta_l$, a parameter of the eigenvector operator $W(\thv)$, is
\begin{equation} \label{eq:CostEigenvectorGradient}
    \begin{split}
        \frac{\partial C_{\rm fsVFF}(U,V, \psi_0)}{\partial \theta_l} = \frac{1}{2} \Big( & C_{\rm fsVFF}(U, W_{l+} D W^\dagger) \,
        - \, C_{\rm fsVFF}(U, W_{l-} D W^\dagger) \\
        + \ & C_{\rm fsVFF}(U, W D (W_{l+})^\dagger) \
        - \  C_{\rm fsVFF}(U, W D (W_{l-})^{\dagger}) \Big) \ .
    \end{split}
\end{equation}
The operator $W_{l+}$ ($W_{l-}$) is generated from the original eigenvector operator $W(\thv)$ by the addition of an extra $\frac{\pi}{2}$ ($-\frac{\pi}{2}$) rotation about a given parameter's rotation axis:
\begin{equation}
    W_{l\pm} := W\left( \thv_{l\pm} \right) \ \ \text{with}  \ \ (\theta_{l\pm})_i := \theta_l \pm \frac{\pi}{2} \delta_{i,l} \; .
\end{equation}

Similarly, the partial derivative with respect to $\gamma_l$, a parameter of the diagonal operator $D(\gav)$, is
\begin{equation} \label{eq:CostDiagonalGradient}
    \begin{split}
        \frac{\partial C_{\rm fsVFF} }{\partial \gamma_l}   =   \frac{1}{n_{\rm eig}}\sum_{k=1}^{n_{\rm eig}} \frac{k}{2} \Big( & C_{\rm fsVFF}^{(k)}\left( U, W  D_{l+} W^\dagger \right)  - C_{\rm fsVFF}^{(k)} \left( U, W  D_{l-} W^\dagger \right) \Big) \, ,
    \end{split}
\end{equation}
where 
\begin{equation}
 C_{\rm fsVFF}^{(k)} :=  1 - | \bra{\psi_0} W D^k W^\dagger U^k \ket{\psi_0} |^2 \, 
\end{equation}
and
\begin{equation}
    D_{l\pm} := D\left( \gav_{l\pm}\right) \ \ \text{with}  \ \ (\gamma_{l\pm})_i := \gamma_l \pm \frac{\pi}{2} \delta_{i,l} \; .
\end{equation}

\medskip

\paragraph*{Derivative with respect to $\gamma_l$.}

Here we provide the derivation of the partial derivative of $C_{\rm fsVFF}(U,V, \psi_0)$ with respect to $\gamma_l$ in \eqref{eq:CostDiagonalGradient}. 
Taking the partial derivative of $ C_{\rm fsVFF}^{(k)}(\vec{\theta},\vec{\gamma})$ with respect to an angle $\gamma_l$ gives
\begin{equation}\label{eq:DifFD}
\begin{aligned} 
     \frac{\partial C_{\rm fsVFF}^{(k)}}{\partial \gamma_l} = &-  \Tr \left[X \left(W \frac{\partial D(k\gav)}{\partial \gamma_l} W^\dagger \right) \dya{Y_k} \left(W D(k\gav)^\dagger W^\dagger \right) \right] \\ 
     &-  \Tr \left[X \left( W D(k\gav) W^\dagger \right) \dya{Y_k} \left( W \frac{\partial D(k\gav)^\dagger}{\partial \gamma_l} W^\dagger \right) \right] \, .
\end{aligned}
\end{equation}
The eigenvector operator, $D$, consists of products of Pauli rotations and can be decomposed as 
\begin{equation}
D = D_L  \exp \left(- \frac{ i k \gamma_l \sigma_l }{2} \right)  D_{R'} \equiv D_L D_R \, ,
\end{equation}
where the operators $D_L$ and $D_{R'}$ consist of all Pauli rotations to the left and right of the $\sigma_l$ rotation respectively and we have defined $D_R = \exp(- i \theta_l \sigma_l/2) D_{R'} $ for convenience. It follows that the differential of $W$ with respect to $\theta_l$ takes the form 
\begin{equation}
\frac{\partial D}{\partial \gamma_l} = - \frac{1}{2} i k D_L \sigma_l D_{R} \; , 
\end{equation}
which on substituting into Eq.~\eqref{eq:DifFD} gives 
\begin{equation}
\begin{aligned}\label{eq:DiagGradInter}
\frac{\partial C_{\rm fsVFF}^{(k)}}{\partial \gamma_l} &= \frac{i k }{2} \bigg( \Tr \left[X W D_L  \sigma_l D_{R} W^\dagger \dya{Y_k} W D^\dagger W^\dagger  \right] - \Tr \left[X  W  D W^\dagger \dya{Y_k} W  D_R^\dagger \sigma_l  D_{L}^\dagger W^\dagger \right] \bigg)  \\
&= \frac{i k }{2}\Tr \left[X  W D_L \left(\sigma_l D_{R} W^\dagger \dya{Y_k} W D_R^\dagger - D_R W^\dagger \dya{Y_k} W  D_R^\dagger \sigma_l \right)  D_{L}^\dagger W^\dagger \right] \\
&= \frac{i k }{2} \Tr \left[X W D_L  [ \sigma_l, \rho_1^{(k)}]  D_{L}^\dagger W^\dagger \right] \, ,
\end{aligned}
\end{equation}
where we have defined
\begin{align}
    &\rho_1^{(k)} = D_{R} W^\dagger \dya{Y_k} W D_R^\dagger \, .
\end{align}
Eq.~\eqref{eq:CostDiagonalGradient} is now obtained directly from Eq.~\eqref{eq:DiagGradInter} via the following identity, which holds for any state $\rho$,
\begin{align}\label{eq:PauliComID}
    i  [ \sigma_l  , \rho] &= e^{ i  \sigma_l \pi /4 } \rho e^{- i  \sigma_l \pi /4 } - e^{ -i  \sigma_l \pi /4 } \rho e^{ i  \sigma_l \pi /4 } \,.
\end{align}
Specifically we find that
\begin{equation}
\begin{aligned}
\frac{\partial C_{\rm fsVFF}^{(k)}}{\partial \gamma_l} = \frac{k}{2}  &\Tr \left[X \otimes W D_L ( e^{ i  \sigma_l \pi /4 }  D_{R} W^\dagger \dya{Y_k} W D_R^\dagger e^{- i  \sigma_l \pi /4 } )  D_{L} W^\dagger  \right] \\
- \frac{k}{2}  &\Tr \left[X  W D_L  ( e^{ -i  \sigma_l \pi /4 }  D_{R} W^\dagger \dya{Y_k} W D_R^\dagger e^{ i  \sigma_l \pi /4 })  D_{L} W^\dagger  \right] \\ 
= \frac{k}{2} &\Tr \left[X \otimes W D_{l-} W^\dagger  \dya{Y_k}  W D_{l-} W^\dagger \right] \\
- \frac{k}{2} &\Tr \left[X \otimes W D_{l+} W^\dagger  \dya{Y_k}  W D_{l+} W^\dagger \right] \,  \\ 
= \frac{k}{2} \Big( & C_{\rm fsVFF}^{(k)}\left( U, W  D_{l+} W^\dagger \right)  - C_{\rm fsVFF}^{(k)} \left( U, W  D_{l-} W^\dagger \right) \Big) \, .
\end{aligned}
\end{equation}
Thus we are left with 
\begin{equation}
    \begin{split}
        \frac{\partial C_{\rm fsVFF}}{\partial \gamma_l}   =   \frac{1}{n_{\rm eig}}  \sum_{k=1}^{n_{\rm eig}} \frac{\partial C_{\rm fsVFF}^{(k)}}{\partial \gamma_l} =  \frac{1}{n_{\rm eig}}\sum_{k=1}^{n_{\rm eig}} \frac{k}{2} \Big( & C_{\rm fsVFF}^{(k)}\left( U, W  D_{l+} W^\dagger \right)  - C_{\rm fsVFF}^{(k)} \left( U, W  D_{l-} W^\dagger \right) \Big) \, . 
    \end{split}
\end{equation}

\paragraph*{Derivative with respect to $\theta_l$.}

Here we provide the derivation of the partial derivative of $C_{\mbox {fsVFF}}(U,V, \psi_0)$ with respect to $\theta_l$ in \eqref{eq:CostEigenvectorGradient}. Taking the partial derivative of $ C_{\rm fsVFF}^{(k)}(\vec{\theta})$ with respect to an angle $\theta_l$ gives 
\begin{equation}\label{eq:DifF}
\begin{aligned} 
     \frac{\partial C_{\rm fsVFF}^{(k)}}{\partial \theta_l} = &-  \Tr \left[X \left(\frac{\partial W}{\partial \theta_l} D W^\dagger \right) \dya{Y_k} \left(W D^\dagger W^\dagger \right) \right] 
     -  \Tr \left[X \left( W D W^\dagger \right) \dya{Y_k} \left(W D^\dagger \frac{\partial W^\dagger}{\partial \theta_l} \right) \right] \\ 
     & -  \Tr \left[X \left( W D \frac{\partial W^\dagger}{\partial \theta_l}\right) \dya{Y_k} \left( W D^\dagger W^\dagger \right) \right] - \Tr \left[X \left(W D W^\dagger \right) \dya{Y_k} \left( \frac{\partial W}{\partial \theta_l} D^\dagger W^\dagger \right) \right] . 
\end{aligned}
\end{equation}
The eigenvector operator, $W$, consists of products of Pauli rotations and can be decomposed as 
\begin{equation}
W = W_L  \exp \left(- \frac{ i \theta_l \sigma_l }{2} \right)  W_{R'} \equiv W_L W_R \, ,
\end{equation}
where the operators $W_L$ and $W_{R'}$ consist of all Pauli rotations to the left and right of the $\sigma_l$ rotation respectively and we have defined $W_R = \exp(- i \theta_l \sigma_l/2) W_{R'} $ for convenience. It follows that the differential of $W$ with respect to $\theta_l$ takes the form 
\begin{equation}
\frac{\partial W}{\partial \theta_l} = - \frac{1}{2} i W_L \sigma_l W_{R} \; , 
\end{equation}
which on substituting into Eq.~\eqref{eq:DifF} gives  \begin{equation}\label{eq:SubDifF}
\begin{aligned}
     \frac{\partial C_{\rm fsVFF}^{(k)}}{\partial \theta_l} 
     = \frac{i}{2} \bigg( \Tr \left[X  W_L  [ \sigma_l  , \rho_2^{(k)}] W_L^\dagger  \right]  
- \Tr \left[X W D  W_R^\dagger [\sigma_l,\rho_3^{(k)}]  W_R D^\dagger W^\dagger  \right] \bigg) \; ,
\end{aligned}
\end{equation}
where we have defined
\begin{align}
    &\rho_2^{(k)} = W_{R} D W^\dagger \dya{Y_k} W D^\dagger W_{R}^\dagger \ \ \ \text{and} \ \ \  \rho_3^{(k)} = W_l^\dagger \dya{Y_k} W_L \ . 
\end{align}
Eq.~\eqref{eq:CostEigenvectorGradient} is now obtained directly from Eq.~\eqref{eq:SubDifF} via Eq.~\eqref{eq:PauliComID}.

\section{Fast forwarding error}\label{ap:errors}

Let us start by writing the state $\ket{\psi_0}$ as
\[
\ket{\psi_0} =\left(\begin{array}{@{}c@{}}
\mathbf{v_{\psi_0}} \\
\hline
 \mathbf{0}
\end{array}\right) 
\] 
to emphasise that $\ket{\psi_0}$ spans a subspace of the total Hilbert space. 
We then note that in the limit in which leakage can be disregarded the evolution unitary can be written in terms of its block decomposition 
\[
U =\left(\begin{array}{@{}c|c@{}}
U_\parallel 
  & \bigzero \\
\hline
  \bigzero &
 U_\perp \\
\end{array}\right) .
\] 
where $U_\parallel$ acts on the subspace spanned by $\psi_0$ and $U_\perp$ acts on the rest of the Hilbert space. 
Let us also write the learnt unitary in a block decomposition form as 
\[
V =\left(\begin{array}{@{}c|c@{}}
V_\parallel 
  & \epsilon_a\\
\hline
  \epsilon_b &
 V_\perp \\
\end{array}\right) .
\] 
In general, the off-diagonal blocks are expected to be close to the null matrices, therefore we can take a perturbative approach to calculating $V^N$.
Expanding $V^N$ to first order in $\epsilon_a$ and $\epsilon_b$ gives
\[
V^N =\left(\begin{array}{@{}c|c@{}}
V_\parallel 
  & \epsilon_a\\
\hline
  \epsilon_b &
 V_\perp \\
\end{array}\right)^N  = 
\left(\begin{array}{@{}c|c@{}}
V_\parallel^N 
  & V_{\epsilon_a}(N)\\
\hline
  V_{\epsilon_b}(N) &
 V_\perp^N \\
\end{array}\right) \, ,
\] 
where 
\begin{equation}
    \begin{aligned}
          &V_{\epsilon_a}(N) := V_\parallel^{N-2} \epsilon_a  V_\parallel + \sum_{k=1}^{N-1} V_\parallel^{N-1 - k } \epsilon_a V_\perp^{k} \\ 
          &V_{\epsilon_b}(N) := V_\perp^{N-2} \epsilon_b  V_\perp + \sum_{k=1}^{N-1} V_\perp^{N-1 - k } \epsilon_b V_\parallel^{k} \, .
    \end{aligned}
\end{equation}
Therefore we can now write 
\[
V^N \ket{\psi_0} = 
\left(\begin{array}{@{}c|c@{}}
V_\parallel^N 
  & V_{\epsilon_a}(N)\\
\hline
  V_{\epsilon_b}(N) &
 V_\perp^N \\
\end{array}\right) \left(\begin{array}{@{}c@{}}
\mathbf{v_{\psi_0}} \\
\hline
 \mathbf{0}
\end{array}\right)  = \left(\begin{array}{@{}c@{}}
V_\parallel^N  \mathbf{v_{\psi_0}} \\
\hline
  V_{\epsilon_b}(N)  \mathbf{v_{\psi_0}}
\end{array}\right)  .
\] 


We are interested in the fast forwarded simulation fidelity 
\begin{equation}
      F_N =  | \bra{\psi_0} {V^\dagger}^N U^N \ket{\psi_0} |^2 \, ,
\end{equation}
which to first order in $\epsilon_a$ and $\epsilon_b$ evaluates to
\begin{equation}
      F_N =  | \bra{\psi_0} {V_\parallel^\dagger}^N U_\parallel^N \ket{\psi_0} |^2 \, . 
\end{equation}
The fidelity between any two quantum states $\ket{\psi_a}$ and $\ket{\psi_b}$ can be related to the trace norm distance between the exact and simulated fast forwarded states via the relation
\begin{equation}
    1 - |\braket{\psi_a | \psi_b }|^2 = \left( \frac{1}{2} || \psi_a -  \psi_b ||_1 \right)^2 \, ,
\end{equation}
where we use the shorthand $\psi_a = \ket{\psi_a} \bra{\psi_a}$ and $\psi_b = \ket{\psi_b} \bra{\psi_b}$. 
Therefore we can write 
\begin{equation}
    1  - F_N =   \left\Vert U_\parallel^N \psi_0 {U_\parallel^\dagger}^N -  V_\parallel^N \psi_0 {V_\parallel^\dagger}^N \right\Vert_1  \, .
\end{equation}
We further note that we can write
\begin{equation}
        \left\Vert U_\parallel^N \psi_0 {U_\parallel^\dagger}^N -  V_\parallel^N \psi_0 {V_\parallel^\dagger}^N \right\Vert_1 =  \left\Vert (U_\parallel^N - V_\parallel^N) \psi_0 {U_\parallel^\dagger}^N +  V_\parallel^N \psi_0 \left({U_\parallel^\dagger}^N - {V_\parallel^\dagger}^N \right)\right\Vert_1 \, ,
\end{equation}
and therefore on applying Holder's inequality, $|| X Y ||_1 \leq || X ||_\infty || Y ||_1$, we have that 
\begin{equation}
    \left\Vert  U_\parallel^N \psi_0 {U_\parallel^\dagger}^N -  V_\parallel^N \psi_0 {V_\parallel^\dagger}^N \right\Vert_1 \leq \left\Vert U_\parallel^N - V_\parallel^N \right\Vert_\infty \left( \left\Vert \psi_0 {U_\parallel^\dagger}^N \right\Vert_1 + \left\Vert \psi_0 {V_\parallel}^N \right\Vert_1 \right) \, . 
\end{equation}
Now from Lemma 1 of \cite{cirstoiu2020variational} we have that  $|| U_\parallel^N - V_\parallel^N ||_\infty  \leq N || U_\parallel - V_\parallel ||_\infty $ and from the unitary invariance of the Schatten norms we have that $|| \psi_0 {U_\parallel^\dagger}^N ||_1  = || \psi_0 {V_\parallel^\dagger}^N ||_1 = || \psi_0 ||_1 = 1$, therefore we are left with
\begin{equation}
  \left\Vert  U_\parallel^N \psi_0 {U_\parallel^\dagger}^N -  V_\parallel^N \psi_0 {V_\parallel^\dagger}^N \right\Vert_1 \leq 2 N \left\Vert U_\parallel - V_\parallel \right\Vert _\infty \, .
\end{equation}
We conclude that the final simulation fidelity is bounded as 
\begin{equation}\label{eq:ErrorBounds}
    1 - F_N \leq  N^2 \left(\left\Vert  U_\parallel - V_\parallel \right\Vert_\infty \right)^2 \, . 
\end{equation}
Thus, the simulation fidelity scales sub-quadratically with the product of the number of simulation steps and the infinity norm distance between the learnt and target unitaries. 

\begin{figure}
\includegraphics[width =0.5\columnwidth]{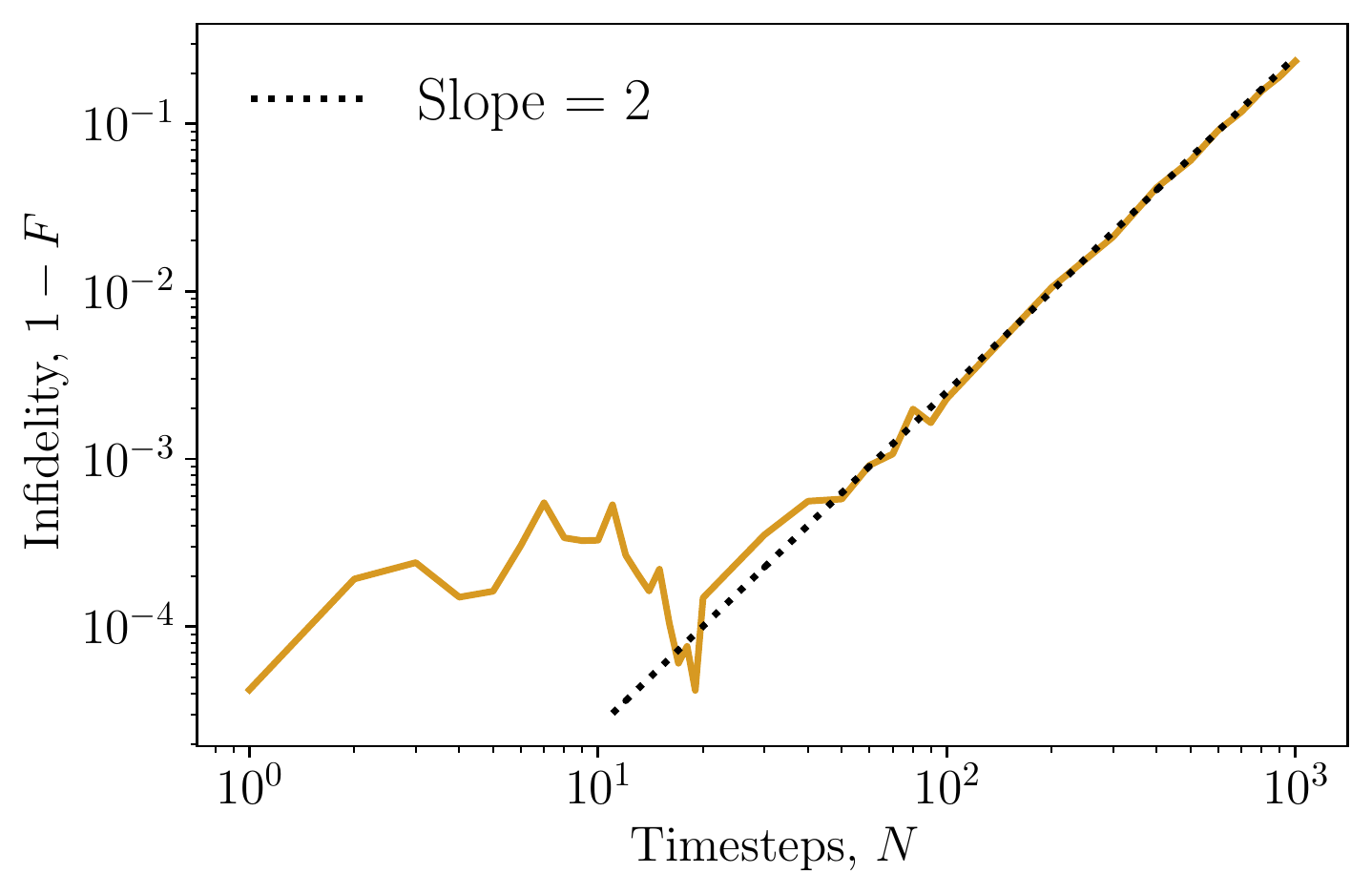}
\vspace{-2mm}
\caption{\textbf{Quadratic Scaling of Infidelity}: The 3-qubit diagonalization used to perform the Eigenvalue Estimation from
Figure~\ref{fig:QEE_eigenvalues} is fast-forwarded with its infidelity, 1 - $F$, plotted against the number of timesteps, $N$. The close alignment of infidelity against the dotted line with form $c \times N^2$ reproduces the scaling of Eq.~\ref{eq:ErrorBounds}.}
\label{fig:quadratic}
\end{figure}

\section{Symmetry preserving adaptive ansatz design}\label{ap:Numerics}


Here we provide further details on the adaptive ansatz used to optimise the noisy simulations and for the batch sampling simulations. This method uses a combination of discrete updates to the structure of the circuit and the training of continuous rotation angles. 
To facilitate discrete structure updates, at the start of the optimization’s inner loop a set of gates initialised to the identity gate are randomly inserted into the ansatz, and when the cost function has plateaued after optimizing the continuous parameters, each gate is tested to see how much the cost function is increased by removing that gate, with a change in cost $\Delta C_1$. In a manner analogous to that used in simulated annealing, if $\text{exp}(-\beta_1\frac{\Delta C_1} {C})$ < rand(0, 1) then the gate is considered not to be significantly contributing to the diagonalization and is deleted from the ansatz. To avoid getting trapped in a local minima, after the cost has plateaued at the end of the inner loop the structure update is accepted if 1-$\text{exp}(-\beta_2\frac{\Delta C_2}{C_{\text{best}}})$ < rand(0, 1), where $\Delta C_2 = C_{\text{plateau}} - C_{\text{best}}$. A subfunction to compile together consecutive identical gates was used to compress the circuit depth.

The Heisenberg class of Hamiltonians are known to observe the particle number-conserving symmetry, so a dictionary of gates $\mathcal{D}$ that all maintain this symmetry was chosen to restrict the size of the subspace the ansatz explored. $\mathcal{D} = \{R_z, R_{zz}, G\}$, where $G$ is an entangling gate conserving the total amplitude within the $\{\ket{01}, \ket{10}\}$ subspace, described in Ref.~\cite{anselmetti2021local}. Linear nearest-neighbour connectivity was enforced during the noisy simulations to avoid the insertion of unwanted SWAP gates. 

In our noiseless batched simulations, the same gate dictionary $\mathcal{D}$ was used as in the noisy simulations, although here all-to-all connectivity was allowed. In a noisy setting, the accumulation of errors due to circuit depth automatically adds a constraint on the circuit depth, however in an unconstrained noise-free setting the insertion of gates may cause the ansatz to arbitrarily grow in length as the optimisation progresses. To prevent this, a regularised cost was used during the structure updates, with a term added to the fsVFF cost to penalise deeper circuits, $C_{\text{reg}} = C_{\text{fsVFF}}*(1 + \frac{N_{gates}}{\lambda})$. The added 1 in the multiplicative factor serves to remove the effect of the regularisation term for small circuits, avoiding early deletion of gates before $C_{\text{fsVFF}}$ has begun to decrease.

\end{document}